\documentclass[twoside]{IEEEtran}

\usepackage{amsmath}
\usepackage{amssymb}

\usepackage{tikz}
\usepackage{float}
\usepackage{multirow}
\usepackage{tabularx}	
\usepackage{pgfplots}
\pgfplotsset{compat=newest}
\pgfplotsset{plot coordinates/math parser=false}
\newlength\figureheight
\newlength\figurewidth

\usepackage{optidef}
\usepackage{algorithm}
\usepackage{algpseudocode}
\usepackage{multicol}

\usepackage{amsthm}
\newtheorem{remark_new}{Remark}
\newtheorem{assumption_new}{Assumption}

\usepackage{comment}

\usepackage{acro}

\DeclareAcronym{lidar}{
short = LiDAR,
long = light detection and ranging sensor,
short-plural = s ,
long-plural = s ,
}

\DeclareAcronym{3D}{
short = 3D ,
long = three-dimensional,
short-plural = s ,
long-plural = s ,
}

\DeclareAcronym{2D}{
short = 2D ,
long = two-dimensional,
short-plural = s ,
long-plural = s ,
}

\DeclareAcronym{ppa}{
short = PPA ,
long = path planning algorithm ,
short-plural = s ,
long-plural = s ,
}

\DeclareAcronym{uas}{
short = UAS ,
long = unmanned aircraft system,
short-plural = s ,
long-plural = s ,
}

\DeclareAcronym{uav}{
short = UAV ,
long = unmanned aerial vehicle,
short-plural = s ,
long-plural = s ,
}

\DeclareAcronym{ecm}{
short = ECM ,
long = energy consumption model,
short-plural = s ,
long-plural = s ,
}

\DeclareAcronym{mpc}{
short = MPC ,
long = model predictive control,
short-plural = s ,
long-plural = s ,
}

\DeclareAcronym{milp}{
short = MILP ,
long = mixed integer linear programming,
short-plural = s ,
long-plural = s ,
}

\DeclareAcronym{lpv}{
short = LPV ,
long = linear parameter-varying,
short-plural = s ,
long-plural = s ,
}

\DeclareAcronym{npv}{
short = NPV ,
long = nonlinear parameter-varying,
short-plural = s ,
long-plural = s ,
}

\DeclareAcronym{ocp}{
short = OCP ,
long = optimal control problem,
short-plural = s ,
long-plural = s ,
}

\DeclareAcronym{bldc}{
short = BLDC,
long = brushless direct current,
short-plural = s ,
long-plural = s ,
}

\DeclareAcronym{dc}{
short = DC,
long = direct current,
short-plural = s ,
long-plural = s ,
}

\DeclareAcronym{esc}{
short = ESC,
long = electric speed controller,
short-plural = s ,
long-plural = s ,
}

\DeclareAcronym{esc-bldc}{
short = ESC-BLDC,
long = ESC-BLDC, 
short-plural = s ,
long-plural = s ,
}
\DeclareAcronym{bldc-esc}{
short = BLDC-ESC,
long = BLDC-ESC, 
short-plural = s ,
long-plural = s ,
}

\DeclareAcronym{lib}{
short = LIB,
long = lithium-ion battery,
short-plural = s ,
long-plural = lithium-ion batteries ,
}

\DeclareAcronym{libc}{
short = LIBC,
long = lithium-ion battery cell,
short-plural = s ,
long-plural = s ,
}

\DeclareAcronym{pwm}{
short = PWM,
long = pulse width modulation,
short-plural = s ,
long-plural = s ,
}

\DeclareAcronym{nps}{
short = NPS,
long = number of polygon sides and affine linear functions,
short-plural = s ,
long-plural = s ,
}

\usepackage[sort&compress,square,comma,numbers]{natbib}

\usepackage[colorlinks=true,urlcolor=blue,citecolor=blue,linkcolor=blue,bookmarks=true]{hyperref}

\begin{document}
\title{A Modular Energy Aware Framework for Multicopter Modeling in Control and Planning Applications}

\author{Sebastian~Gasche, Christian~Kallies, Andreas~Himmel, and Rolf~Findeisen\thanks{S. Gasche and C. Kallies are with the Institute of Flight Guidance, German Aerospace Center, Brunswick, Germany.}\thanks{S. Gasche, A. Himmel and R. Findeisen are with the Technical University of Darmstadt, Darmstadt, Germany.}}

\markboth{PREPRINT: This work has been submitted to the IEEE for possible publication. Copyright may be transferred without notice.}%
{S. Gasche \MakeLowercase{\textit{et al.}}: A Modular Energy Aware Framework for Multicopter Modeling in Control and Planning Applications}

\maketitle

\begin{abstract}
Unmanned aerial vehicles (UAVs), especially multicopters, have recently gained popularity for use in surveillance, monitoring, inspection, and search and rescue missions. Their maneuverability and ability to operate in confined spaces make them particularly useful in cluttered environments. For advanced control and mission planning applications, accurate and resource-efficient modeling of UAVs and their capabilities is essential. This study presents a modular approach to multicopter modeling that considers vehicle dynamics, energy consumption, and sensor integration. The power train model includes detailed descriptions of key components such as the lithium-ion battery, electronic speed controllers, and brushless DC motors. Their models are validated with real test flight data. In addition, sensor models, including LiDAR and cameras, are integrated to describe the equipment often used in surveillance and monitoring missions. The individual models are combined into an energy-aware multicopter model, which provide the basis for a companion study on path planning for unmanned aircaft system (UAS) swarms performing search and rescue missions in cluttered and dynamic environments. The flexible modeling approach enables easy description of different UAVs in a heterogeneous UAS swarm, allowing for energy-efficient operations and autonomous decision making for a reliable mission performance.
\end{abstract}


\IEEEpeerreviewmaketitle
\section{Introduction}
\label{cha:Intro}
Today, \acp{uav} are used in various industrial and civilian applications, including search and rescue, surveillance, and inspection missions, due to their flexibility, ease of deployment, and ability to access hard-to-reach areas. They are available in a wide range of designs, such as fixed-wing, rotary-wing, and hybrid configurations, each suited to specific mission profiles \citep{Singhal2018}. Among them, multicopters are notable for their high maneuverability, stability, and ability to hover, making them particularly effective in cluttered environments such as urban, forested, or disaster-affected areas. Motivated by their suitability for tasks in confined and complex environments where other types of \acp{uav} may struggle, this study focuses on multicopters \cite{Mohsan2023}.

Many tasks such as mission planning, path planning, control, and state estimation require accurate models of the \ac{uav}'s dynamics to ensure that the \ac{uav} performs as expected. This includes the \ac{uav}'s vehicle dynamics, such as forces and moments acting on the body, as well as rotational and translational motions. In addition, its interaction with the environment, including disturbances such as wind or turbulence, must also be taken into account. Besides the vehicle dynamics, the dynamics of its energy consumption is often important to model. As the energy consumption has a direct influence on the flight time, an accurate modeling enables to realistically estimate the flight endurance and the mission feasibility.

These models are central to several control and optimization methods, such as \ac{mpc}, which require accurate models of the system to predict future states and adjust controls accordingly \citep{Wei2022,Malyuta2022,Quirynen2024}. For instance, in our companion study \cite{Gasche2024_PPA} on path planning for a heterogeneous \ac{uas} swarm with the goal of planning flight paths for search and rescue missions, models of the \acp{uav}' vehicle dynamics and energy consumption are used. The path planning algorithm is based on \ac{mpc} and \ac{milp}, which means that a mathematical model of the vehicle, the environment, and the mission goal is needed for path planning and decision making. Based on these models, the future behavior of the \acp{uas} is predicted and optimized in terms of mission success, energy efficiency, and safety, enabling cooperative and sustainable swarm guidance. Furthermore, the estimation of the remaining energy capacity of the \acp{uav}, which is integrated into the decision making, enables the autonomous return of discharged \acp{uav} to their landing sites for recharging. Other path planning algorithms, such as rapidly-exploring random trees \citep{Kuffner2000,kingston2018} and genetic algorithms \citep{Patle2019,Mac2016}, also benefit from such models to incorporate motion constraints during sampling.

Sensors, such as cameras and LIDARs, which are common payloads for \ac{uas} missions, also impose constraints on \ac{uas} operations to ensure valid data collection. Therefore, the effects of the sensors on the dynamics and energy consumption of the \ac{uav} should be considered for accurate planning tasks.

In modeling, there are three primary types of models that differ in required knowledge of the inner dynamics of the system, required data on the system response, and descriptiveness.
\textit{White-box models}, also known as physics-based or first-principles models, analytically model system behavior based on physical equations where all parameters must be known. While highly interpretable and providing detailed insight into underlying physics, they require precise knowledge of all system components, challenging for complex systems. They offer high accuracy when the system is well understood but can be computationally expensive, limiting their use in real-time applications.
\textit{Black-box models} rely purely on input-output data without requiring detailed understanding of inner system dynamics. These models use machine learning or system identification methods to model system behavior as a relation between input and output. They can capture complex nonlinear system dynamics and are easy to implement with large datasets but lack physical interpretability. Additionally, their limited extrapolation capabilities result in inaccuracies when encountering unlearned situations, as accuracy depends on training data quantity and quality.
\textit{Grey-box models} combine elements of white-box and black-box models, using partial knowledge of inner dynamics while estimating unknown parameters or component dynamics from data. This offers a balance between accuracy and computational efficiency, particularly useful in practical applications, where some system knowledge is available but precise modeling of every component or process is impractical. However, determining the optimal model structure can be challenging, requiring both physical knowledge and data \citep{Ali2023,vivian2024}.

In this study, a grey-box modeling approach is used for its balance of accuracy, flexibility, and computational efficiency. It allows the incorporation of known \ac{uav} dynamics \citep{garcia2005,cai2011,nonami2010,Wall2020}, the approximation of difficult-to-model dynamics, and the consideration of uncertain or unknown parameters. In addition, we are aiming at a modular model for multicopters, which offers significant advantages in terms of flexibility and configurability. Multicopters can vary widely in the number of rotors, payloads, sensors, and battery capacities. A modular approach allows components such as motors, propellers, batteries and sensors to be easily reconfigured or replaced without overhauling the entire model. This is particularly useful when planning missions where the \ac{uav} may need to be configured to perform different tasks or operate in different environments. For example, by swapping out modules such as batteries or motors, the model can be adapted for missions that require heavier payloads or those that prioritize longer endurance. 

\acp{uav} are typically driven by electric motors and propelled by electrical energy stored in a battery. As their size increases, there are also \acp{uav} that are propelled by fossil fuels in combination with combustion engines or jet engines. However, only electric-propelled \acp{uav} are considered in this study. Numerous modeling approaches have been proposed to describe the energy consumption of electric-propelled \acp{uav}, ranging from high-level empirical models to detailed physics-based models.\citet{Zhang(2021)} and \citet{Muli2022} review \acp{ecm} of electric-propelled \acp{uav} and classify them into integrated models, regression models, and component models.
\textit{Integrated models} combine various aerodynamic and design aspects into a single critical parameter, the lift-to-drag ratio, to represent energy efficiency. For instance, \citet{Andrea2014} integrated model introduces this approach by considering the mass, the velocity, the lift-drag-ratio, the power train efficiency, and the power consumption of the avionics, providing a broad yet cohesive estimation of energy usage for \acp{uav} across different flight phases. This method is efficient for high-level planning, though it highly depend on the choice of the lift-to-drag ratio and neglects detailed forces, which limits the model to a specific operation case.  
\textit{Component models} decompose energy consumption into separate segments, such as hovering, takeoff, landing, and cruising, to estimate energy more granularly. This approach allows detailed representations of energy requirements by considering individual forces, such as the aircraft’s weight force and various drag forces.  \citet{Stolaroff2018} apply a two-component model that includes the thrust required to compensate for weight and to counteract parasite drag. While this model can reflect variations in power demands across different phases of a flight, it may be complex to parameterize accurately and often require substantial empirical data for calibration. 
\textit{Regression models} rely on empirical data from field tests, such as the work of \citet{Alyassi2023}, who utilize nonlinear regression with multiple variables, including payload mass, velocity, acceleration and wind conditions, to predict energy consumption. These models are useful in capturing energy requirements in real-world settings, especially where environmental factors significantly impact performance. By fitting data to real-world conditions, regression models can produce realistic estimations for specific \acp{uav} and operational parameters. However, they are limited by the data available and may not generalize well to different \ac{uav} designs or operation scenarios \citep{Zhang(2021),Muli2022}.

\citet{Asti2020} propose a different approach to develop an energy-efficient obstacle avoidance. Here the change of kinetic and potential energy is used to evaluate the efficiency of a maneuver. This approach is simple and does not need further insight into the \ac{uav} design besides the total mass of the \ac{uav}. While it can evaluate the efficiency of a maneuver, it is not suitable for accurately estimating energy consumption. 

In contrast to the mentioned models, we adopt a modular component-based approach, where the energy consumption is modeled by considering key components of the power train such as the \ac{lib}, \acp{esc}, \ac{bldc} motors and rotors. This level of abstraction allows for greater accuracy and flexibility, allowing the model to be adapted for different \ac{uav} configurations and use cases. Combined with an accurate model of the \ac{uav} dynamics, this model can be adapted to meet our requirements for accuracy, generalizability, and computational efficiency. By modeling the individual components, not only specific maneuvers are considered, but rather a variety of maneuvers. This allows for dynamic operations without having to discretize into different flight phases. Meanwhile, we aim to use common data sheet information, while only relying on few test flights to identify or calibrate model parameters.

This study is structured as follows: Section \ref{cha:UAVModel} details the modeling of multicopter \acp{uav}, covering the vehicle's dynamics and physical properties. Section \ref{cha:ECM} presents the modeling of the energy consumption of electric-propelled \acp{uav}. Section \ref{cha:sensor} introduces the sensor models, specifically the camera and LiDAR, and their impact on the \ac{uav}'s performance. In Section \ref{cha:model}, these models are combined to be used in advanced mission planning and control applications. Section \ref{cha:Discussion} discusses the validation of the energy consumption model and examines the uncertainties introduced by environmental factors. Finally, Section \ref{cha:Conclusion} provides conclusions and suggests future work based on the outcomes of this study.

\section{Unmanned Aerial Vehicle Modeling}
\label{cha:UAVModel}
In the following, we look at rotary-wing \acp{uav} (multicopters), which use motors with attached rotors to generate a downward thrust to take off or remain in flight. Their high maneuverability (capable to hover and fly at high or low speeds) makes them ideal for surveillance or monitoring missions.
Multicopters have a multiplicity of arms, each equipped with a motor driving a fixed rotor. The rotors rotate either clockwise or counterclockwise in an alternating pattern to balance the system with regard to the drag moment generated by the rotors in stationary flight. The following mathematical formulations of the multicopter's \textit{kinematics} and \textit{dynamics} are based on \citep{Elkholy(2014),Lovas(2018),Nagaty(2013),Wang(2016),Osmić(2016)}. To derive the multicopter model we assume:
\begin{assumption_new}
\label{ass:UAV_1}
The multicopter is axis-symmetric with respect to the body-fixed frame and it's body is nearly spherical.
\end{assumption_new}
\begin{assumption_new}
\label{ass:UAV_2}
The multicopter's actuators, like the \ac{bldc} motors, \acp{esc} and rotors are identical.
\end{assumption_new}

\vspace{-0.5cm}
\begin{figure}[H]
\centering
\input{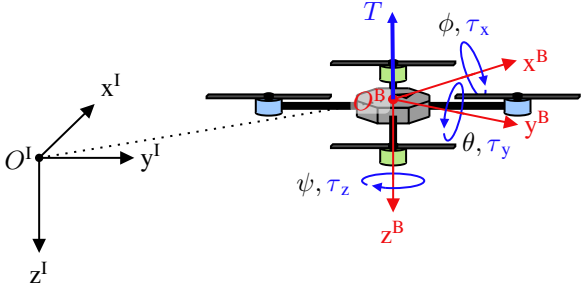}
\vspace{-0.5cm}
\caption{Frames of reference (black: inertial frame, red: body-fixed frame); Forces and torques acting on the body's center of mass (blue)}
\label{fig:QC_Frames}
\end{figure}
\vspace{-0.5cm}

\subsection{Multicopter Kinematics}
We define two frames of reference, shown in Fig.~\ref{fig:QC_Frames}. The inertial frame of reference is an earth-fixed north-east-down frame and its origin \(O^\text{I}\) is attached to the earth's surface. The body-fixed frame is a forward-right-down frame and its origin \(O^\text{B}\) is attached to the multicopter's center of mass. Both frames of reference are right-handed coordinate systems. 

The multicopter is a six-degree-of-freedom (6 DOF) underactuated system, meaning that the rotational system is fully actuated, meanwhile, the translational system is underactuated. The position \(\mathbf{P}^\text{I} = \left(x,y,z\right)^\top\) of the multicopter represents the distance between the origins of the reference frames and is defined in the inertial frame. The orientation  \(\mathbf{\Psi} =(\phi,\theta,\psi)^\top\) of the multicopter is represented by Euler angles, also known as yaw angle \(\psi\) (rotation around the z-axis), pitch angle \(\theta\) (rotation around the y-axis), and roll angle \(\phi\) (rotation around the x-axis). It is defined as the rotation between the inertial and body-fixed frame. 
Since some values are measured in the inertial frame and others in the body-fixed frame, we define two transformation matrices to convert values from the body-fixed frame to the inertial frame. 
A vector, defined in the inertial frame, is obtained by the product of the corresponding vector in the body-fixed frame and the rotation matrix 
\begingroup
\scriptsize
\begin{equation*}
    \mathbf{R}_\text{B}^\text{I} = 
    \begin{bmatrix}
    \text{c}(\psi)\text{c}(\theta) & \text{c}(\psi)\text{s}(\theta)\text{s}(\phi)-\text{s}(\psi)\text{c}(\phi) & \text{c}(\psi)\text{s}(\theta)\text{c}(\phi)+\text{s}(\psi)\text{s}(\phi) \\
    \text{s}(\psi)\text{c}(\theta) & \text{s}(\psi)\text{s}(\theta)\text{s}(\phi)+\text{c}(\psi)\text{c}(\phi) & \text{s}(\psi)\text{s}(\theta)\text{c}(\phi)-\text{c}(\psi)\text{s}(\phi) \\
    -\text{s}(\theta) & \text{c}(\theta)\text{s}(\phi) & \text{c}(\theta)\text{c}(\phi) 
    \end{bmatrix},
\end{equation*}
\endgroup
where \(\text{s},\text{c},\text{t}\) are abbreviations for \(\sin,\cos,\tan\).
Accordingly, the translational velocity \(\dot{\mathbf{P}}^\text{I} = (v_\text{x},v_\text{y},v_\text{z})^\top\) is obtained by 
\begin{equation*}
\dot{\mathbf{P}}^\text{I} = \mathbf{R}_\text{B}^\text{I}\dot{\mathbf{P}}^\text{B},
\end{equation*}
where \(\dot{\mathbf{P}}^\text{B}\) is the velocity vector in the body-fixed frame.
Likewise, the euler rates \(\dot{\mathbf{\Psi}} =(\dot{\phi},\dot{\theta},\dot{\psi})^\top\) are obtained by 
\begin{equation}
\label{eqn:euler_rates}
\dot{\mathbf{\Psi}} = \mathbf{R}_\Psi\,\boldsymbol{\omega}^\text{B}.
\end{equation}
Here, the angular velocity \(\omega^\text{B} = (\omega_\text{x},\omega_\text{y},\omega_\text{z})^\top\) in the body-fixed frame is transformed by the angular transformation matrix
\begin{equation*}
\label{eqn:TMatrix_rot}
    \mathbf{R}_\Psi = 
    \begin{bmatrix}
    1 & \text{s}(\phi)\text{t}(\theta) & \text{c}(\phi)\text{t}(\theta) \\
    0 & \text{c}(\phi)  & -\text{s}(\phi) \\
    0 & \text{s}(\phi)/\text{c}(\theta)  & \text{c}(\phi)/\text{c}(\theta)
    \end{bmatrix}.
\end{equation*}
\begin{remark_new}
\label{remark:UAV_1}
    We constrain the Euler angles \(\phi,\theta \in \left(-\frac{\pi}{2},\frac{\pi}{2}\right)\) to avoid singularities in \(\mathbf{R}_\Psi\). This assumption is feasible if the multicopter does not perform aggressive maneuvers \citep{Fouad(2017),Osmić(2016)}. 
\end{remark_new}
\subsection{Multicopter Dynamics}
The motion of the multicopter is divided into a rotational and a translational motion component. To control the rotational motion, we apply torque to the multicopter's center of mass. Fig.~\ref{fig:QC_Frames} shows the controllable torques \(\boldsymbol{\tau}^\text{B} = (\tau_\text{x},\tau_\text{y},\tau_\text{z})^\top\), as well as the combined thrust \(T\) of all rotors, which is used to control the translational motion.

The rotational equations of motion in the body-fixed frame derive from the Newton-Euler formalism 
\begin{equation}
\label{eqn:NewtonEuler}
    \boldsymbol{\tau}^\text{B} = \mathbf{J}\dot{\boldsymbol{\omega}}^\text{B} + \boldsymbol{\omega}^\text{B} \times \mathbf{J}\,\boldsymbol{\omega}^\text{B} + \boldsymbol{\tau}_\text{G}^\text{B} + \boldsymbol{\tau}_\text{D}^\text{B},
\end{equation}
where \(\dot{\boldsymbol{\omega}}^\text{B} = \left(\dot{\omega}_\text{x},\dot{\omega}_\text{y},\dot{\omega}_\text{z}\right)^\top\) is the angular acceleration. Due to Assumption \ref{ass:UAV_1}, the inertia tensor \(\mathbf{J} = \text{diag}\{J_\text{xx}, J_\text{yy}, J_\text{zz}\}\) has only entries on the diagonal representing the moments of inertia around the body axes. The gyroscopic effect of the angular motion of the rotors is considered by
\begin{equation*}
\label{eqn:MG}
    \boldsymbol{\tau}_\text{G}^\text{B} = \boldsymbol{\omega}^\text{B} \times
    \begin{pmatrix}0 \\ 0 \\ J_\text{r}\,\Omega_\text{r}\end{pmatrix},
\end{equation*}
where \(J_\text{r}\) and \(\Omega_\text{r}\) are the inertia moment of a rotor and the difference in rotor speeds, respectively. The drag torque
\begin{equation*}
    \boldsymbol{\tau}_\text{D}^\text{B} = \mathbf{D}_\tau\boldsymbol{\omega}^\text{B}
\end{equation*}
accounts for the air drag, which is approximately proportional to the angular velocity \(\omega^\text{B}\) and depends on the angular air resistance coefficients within matrix \(\mathbf{D}_\tau = \text{diag}\{c_{\tau\text{x}},c_{\tau\text{y}},c_{\tau\text{z}}\}\). 

The translational equations of motion in the inertial frame derive from Newton's second law 
\begin{equation}
\label{eqn:Newton2th}
m\,\ddot{\mathbf{P}}^\text{I} = \mathbf{F}_\text{G}^\text{I} + \mathbf{R}_\text{B}^\text{I}\,\mathbf{F}^\text{B} - \mathbf{F}^\text{I}_\text{D},
\end{equation}
where \(\ddot{\mathbf{P}}^\text{I}=\left(\ddot{x},\ddot{y},\ddot{z}\right)^\top\) is the translational acceleration. 
The gravitational force 
\[\mathbf{F}_\text{G}^\text{I} = \left(0,0,m\,\text{g}\right)^\top,\]
depends on the total mass of the body \(m\) as well as the acceleration of free fall \(\text{g}\), while the non-gravitational force 
\[\mathbf{F}^\text{B} = \left(0,0,-T\right)^\top,\]
results from the thrust \(T\) of all rotors. 
Lastly, the drag force
\begin{equation*}
\mathbf{F}_\text{D}^\text{I} = 
\mathbf{D}_\text{F}\mathbf{v}_\text{a}^\text{I}
\end{equation*}
accounts for the air drag, which is approximately proportional to the air velocity \(\mathbf{v}_\text{a}^\text{I}\) and further depends on the resistance coefficients within the matrix \(\mathbf{D}_\text{F} = \text{diag}\{c_\text{Fx},c_\text{Fy},c_\text{Fz}\}\) \cite{Hattenberger2023}. Here, the air velocity \(\mathbf{v}_\text{a}^\text{I} = \dot{\mathbf{P}}^\text{I} - \mathbf{v}_\text{w}^\text{I}\) relates the \ac{uav}'s translational velocity \(\dot{\mathbf{P}}^\text{I}\) to the wind velocity \(\mathbf{v}_\text{w}^\text{I}=(v_\text{w,x},v_\text{w,y},v_\text{w,z})^\top\). 

\subsection{Multicopter Models}
\label{sec:Nonlin_UAV}
We rearrange and combine \eqref{eqn:euler_rates}, \eqref{eqn:NewtonEuler} and \eqref{eqn:Newton2th} to obtain the general nonlinear state space multicopter model 
\begin{equation}
\label{eqn:Nonlin_Multicopter}
    \dot{\mathbf{x}}_\text{u} = \mathbf{f}_\text{u}(\mathbf{x}_\text{u},\mathbf{u}_\text{u},\mathbf{d}_\text{u}) + \boldsymbol{\Gamma}_\text{u,x},
\end{equation}
with the state \(\mathbf{x}_\text{u} = (x,y,z,v_\text{x},v_\text{y},v_\text{z},\phi,\theta,\psi,\omega_\text{x},\omega_\text{y},\omega_\text{z})^\top\), the input \(\mathbf{u}_\text{u} = (T,\tau_\text{x},\tau_\text{y},\tau_\text{z},\Omega_\text{r})^\top\) and the external disturbance \(\mathbf{d}_\text{u} = (v_\text{w,x},v_\text{w,y},v_\text{w,z})^\top\). Here, the right-hand side reads
\begin{align*}
    \mathbf{f}_\text{u}(\mathbf{x}_\text{u},\mathbf{u}_\text{u},\mathbf{d}_\text{u}) &=
    \begin{pmatrix}  
    \dot{\mathbf{P}}^\text{I}\\
    \frac{1}{m}\left(\mathbf{F}_\text{G}^\text{I} 
    + \mathbf{R}_\text{B}^\text{I}\,\mathbf{F}^\text{B} - \mathbf{F}_\text{D}^\text{I}\right) \\
    \mathbf{R}_\Psi\,\boldsymbol{\omega}^\text{B} \\
    \mathbf{J}^{-1}\left(\boldsymbol{\tau}^\text{B} - \boldsymbol{\omega}^\text{B} \times \mathbf{J}\,\boldsymbol{\omega}^\text{B} - \boldsymbol{\tau}_\text{G}^\text{B}- \boldsymbol{\tau}_\text{D}^\text{B}\right)
    \end{pmatrix}.
\end{align*}
Additionally, \(\boldsymbol{\Gamma}_\text{u,x}\) represents unknown uncertainties, resulting from modeling inaccuracies and turbulences.

As the path planner in \cite{Gasche2024_PPA}, many real-time applications require discrete-time linear models. Therefore, we derive in Appendix \ref{cha:AppendixD}, the corresponding multicopter model 
\begin{equation}
\label{eqn:Lin_Multicopter}
    \mathbf{x}_\text{u}(k + 1) = \mathbf{A}_\text{d,u}\,\mathbf{x}_\text{u}(k)+\mathbf{B}_\text{d,u}\,\tilde{\mathbf{u}}_\text{u}(k)+\mathbf{H}_\text{d,u}\,\mathbf{d}_\text{u}(k) + \boldsymbol{\Gamma}_\text{u,x}(k),
\end{equation}    
where \(\mathbf{A}_\text{d,u}\), \(\mathbf{B}_\text{d,u}\) and \(\mathbf{H}_\text{d,u}\) are the discrete-time system, input and disturbance matrices. For the chosen set point, the hovering state \(\mathbf{x}_\text{u,SP} = (0,\dots,0)^\top\) without any external disturbance \linebreak \(\mathbf{d}_\text{u,SP} = (0,\dots,0)^\top\), the multicopter maintains its position and the thrust \(T_\text{SP}=m\,\text{g}\) compensates for the weight force. Meanwhile, the input is reduced to \[\tilde{\mathbf{u}}_\text{u}=(L,\tau_x,\tau_y,\tau_z)^\top,\] where the lift \(L\) is the thrust component acting in negative \(\text{z}^\text{I}\)-direction, which is added to the hovering thrust \(T_\text{SP}\). 

\subsection{Adapting For Specific Multicopter Configurations}
\label{sec:UAV_config}
\begin{figure}[H]
\centering
\input{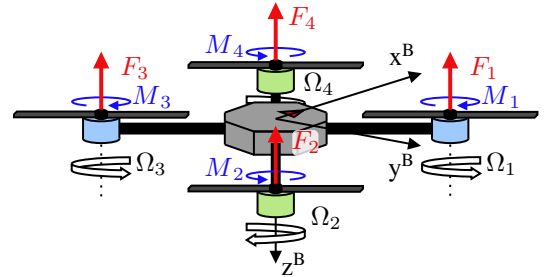}
\vspace{-0.25cm}
\caption{Aerodynamic forces and torques of a quadcopter in X-configuration}
\label{fig:QC_Forces}
\end{figure}

The various multicopter configurations differ in their positioning and number of rotors. Each rotor is fixed to a motor, rotating with the motor speed \(\Omega_i, \ i \in \{1,\dots,N_\text{M}\}\), where \(N_\text{M}\) is the number of motors. 
To adapt the generalized multicopter model for a specific multicopter configuration, the input has to be defined depending on these motor speeds. In the following, we define clockwise rotations as positive and counterclockwise rotations as negative. Accordingly, the difference in rotor speeds is given by
\begin{equation}
\label{eqn:Omega_r}
    \Omega_\text{r} = \sum_{i=1}^{N_\text{M}} \text{sign}(\Omega_i)\,\Omega_i.
\end{equation}
Each rotor generates an upwards-pointing aerodynamic force 
\begin{equation}
\label{eqn:Fi}
    F_i =  k_\text{F}\,\Omega_i^2, \quad \forall \, i \in \{1,\dots,N_\text{M}\},
\end{equation}
and a rotation-counteracting aerodynamic drag torque 
\begin{equation}
\label{eqn:Mi}
    M_i =  k_\text{M}\,\Omega_i^2, \quad \forall \, i \in \{1,\dots,N_\text{M}\},
\end{equation}
where \(k_\text{F}\) and \(k_\text{M}\) are the aerodynamic force and torque constants and Assumption \ref{ass:aero_const} applies \cite{Elkholy(2014),Lovas(2018),Nagaty(2013)}. \begin{assumption_new}
\label{ass:aero_const}
    For simplicity, we assume that the aerodynamic parameters \(k_\text{F}\) and \(k_\text{M}\) are constant. However, in reality, they depend on the rotor speed, airflow, and air pressure. 
\end{assumption_new}
According to Fig.~\ref{fig:QC_Forces}, the thrust 
\begin{equation}
\label{eqn:T}
    T = \sum_{i=1}^{N_\text{M}} F_i,
\end{equation}
combines the \(N_\text{M}\) forces, defined by \eqref{eqn:Fi}. 
Meanwhile, the torques are given by 
\begin{equation}
\label{eqn:tau}     
    \tau_\text{x} = \sum_{i=1}^{N_\text{M}} -l_{\text{y},i}\,F_i, 
    \ \    
    \tau_\text{y} = \sum_{i=1}^{N_\text{M}} l_{\text{x},i}\,F_i,  
    \ \ 
    \tau_\text{z} = \sum_{i=1}^{N_\text{M}} -\text{sign}(\Omega_i)\,M_i,
\end{equation}
where \(\mathbf{l}_{i}^\text{B} = (l_{\text{x},i},l_{\text{y},i})^\top\) indicates the  \(i^\text{th}\) rotor's position. 
\vspace{-0.25cm}
\vspace{-0.5cm}

\subsection{Vehicle Capabilities}
\label{sec:UAV_constr}
In order to consider the vehicle’s limits, the model is constrained. It is common practice to divide the velocity of a multicopter into two parts: the ground velocity 
\begin{equation*}
\label{constr:ground_velocity}
v_\text{g} = \sqrt{v_\text{x}^2 + v_\text{y}^2}, \quad
\text{s.t.} \
v_\text{g} \leq v_\text{g,max}
\end{equation*}
and the climb/descent velocity
\begin{equation*}
\label{constr:climb_velocity}
v_\text{c} = |v_\text{z}|, \quad
\text{s.t.} \
v_\text{c} \leq v_\text{c,max},
\end{equation*}
which are constrained by their respective maximum values \(v_\text{g,max}\) and \(v_\text{c,max}\).
Moreover, the tilt angle 
\begin{equation*}
\label{constr:tilt}
\begin{split}
&\alpha = \cos^{-1}(\cos(\phi)\cos(\theta)) \approx \sqrt{\phi^2+\theta^2},  \\ 
&\text{s.t.} \
\alpha \leq \alpha_\text{max},
\end{split}
\end{equation*}
is constrained by its maximum value \(\alpha_\text{max}\) and Remark \ref{remark:UAV_1} must be considered to avoid singularities. Constraints on the angular rates \(\omega^\text{B}\) should be included if the multicopter is equipped with sensible instruments. 

\noindent
To represent the motors’ capabilities, the input is constrained using \eqref{eqn:T}, \eqref{eqn:tau} and the motor speed limits
\begin{equation*}
\label{constr:motor_speeds}
0 \leq \Omega_i \leq \Omega_\text{max}, \ \ \forall \, i \in \{1,\dots,N_\text{M}\}.
\end{equation*}

\begin{remark_new}
\label{remark:psi}
The approximation error of the linearized model decreases if the yaw angle \(\psi\) is constrained to \(\psi\approx 0\).
\end{remark_new}

\section{Energy Consumption Modeling}
\label{cha:ECM}
In this study, we employ a component modeling approach for the energy consumption modeling as the detailed model design enables us to consider different \ac{uav} designs, environmental influences, and use cases. In combination with an accurate \ac{uav} model, a component model is adapted to fulfill our requirements of accuracy, generalizability, and simulation resource demand. In the following, we derive the individual components of the power train and combine them to obtain the \ac{ecm} for electric-propelled \acp{uav}. 

\begin{figure}[H]
\centering
\input{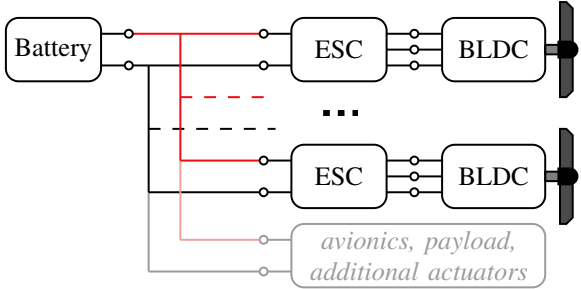}
\caption{Common power train of an electric-propelled \ac{uav}}
\label{fig:EC_0}
\end{figure}

Fig.~\ref{fig:EC_0} shows the power train of an electric-propelled \ac{uav}, which is divided into three prominent components. The \acf{lib} stores the energy, which the \acf{esc} transfers to the \acf{bldc} motor, which drives a fixed rotor. The \ac{esc} controls the \ac{bldc} motor, depending on a \ac{pwm} control command provided by the flight controller. The component model could be extended by including the avionics, the payload, or additional actuators. However, the complexity increase is only meaningful if they consume significant energy compared to the \ac{bldc} motors.

\subsection{Lithium-Ion Battery}
\label{sec:Battery}
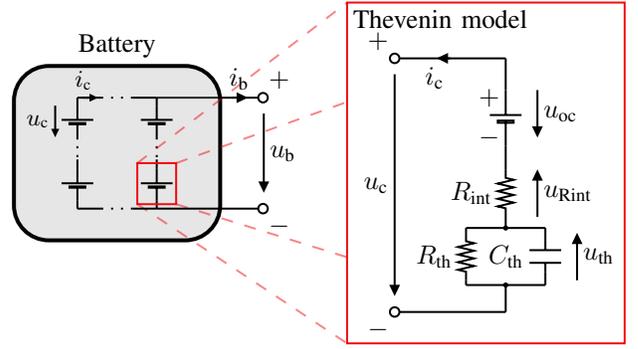
\begin{figure}[H]
\centering
\tikzset{every picture/.style={line width=0.75pt}} %

\begin{tikzpicture}[x=0.6pt,y=0.6pt,yscale=-1,xscale=1]

\draw    (100,65) -- (140,65) ;
\draw  [fill={rgb, 255:red, 155; green, 155; blue, 155 }  ,fill opacity=0.25 ][line width=1.5]  (10,67) .. controls (10,54.85) and (19.85,45) .. (32,45) -- (118,45) .. controls (130.15,45) and (140,54.85) .. (140,67) -- (140,133) .. controls (140,145.15) and (130.15,155) .. (118,155) -- (32,155) .. controls (19.85,155) and (10,145.15) .. (10,133) -- cycle ;
\draw    (50,65) -- (65,65) ;
\draw    (53,65) -- (60,65.01) ;
\draw [shift={(63,65.02)}, rotate = 180.09] [fill={rgb, 255:red, 0; green, 0; blue, 0 }  ][line width=0.08]  [draw opacity=0] (5.36,-2.57) -- (0,0) -- (5.36,2.57) -- cycle    ;
\draw    (36,70) -- (36,87.01) ;
\draw [shift={(36,90.01)}, rotate = 270] [fill={rgb, 255:red, 0; green, 0; blue, 0 }  ][line width=0.08]  [draw opacity=0] (5.36,-2.57) -- (0,0) -- (5.36,2.57) -- cycle    ;
\draw   (50,130) -- (50,121) (40,119) -- (60,119) (50,119) -- (50,110) (45,121.8) -- (45,121) -- (55,121) -- (55,121.8) -- (45,121.8) -- cycle ;
\draw   (50,90) -- (50,81) (40,79) -- (60,79) (50,79) -- (50,70) (45,81.8) -- (45,81) -- (55,81) -- (55,81.8) -- (45,81.8) -- cycle ;
\draw   (100,90) -- (100,81) (90,79) -- (110,79) (100,79) -- (100,70) (95,81.8) -- (95,81) -- (105,81) -- (105,81.8) -- (95,81.8) -- cycle ;
\draw   (100,130) -- (100,121) (90,119) -- (110,119) (100,119) -- (100,110) (95,121.8) -- (95,121) -- (105,121) -- (105,121.8) -- (95,121.8) -- cycle ;
\draw    (50,135) -- (65,135) ;
\draw  [dash pattern={on 0.75pt off 3pt}]  (65,135) -- (85,135) ;
\draw    (147.5,64.99) -- (154.5,65) ;
\draw [shift={(157.5,65)}, rotate = 180.08] [fill={rgb, 255:red, 0; green, 0; blue, 0 }  ][line width=0.08]  [draw opacity=0] (7.14,-3.43) -- (0,0) -- (7.14,3.43) -- cycle    ;
\draw   (164.02,135) .. controls (164.02,133.34) and (165.36,132) .. (167.02,132) .. controls (168.68,132) and (170.02,133.34) .. (170.02,135) .. controls (170.02,136.66) and (168.68,138) .. (167.02,138) .. controls (165.36,138) and (164.02,136.66) .. (164.02,135) -- cycle ;
\draw   (164.02,65) .. controls (164.02,63.34) and (165.36,62) .. (167.02,62) .. controls (168.68,62) and (170.02,63.34) .. (170.02,65) .. controls (170.02,66.66) and (168.68,68) .. (167.02,68) .. controls (165.36,68) and (164.02,66.66) .. (164.02,65) -- cycle ;
\draw    (85,65) -- (100,65) ;
\draw  [dash pattern={on 0.75pt off 3pt}]  (65,65) -- (85,65) ;
\draw    (50,65) -- (50,70) ;
\draw    (100,65) -- (100,70) ;
\draw    (50,130) -- (50,135) ;
\draw    (100,130) -- (100,135) ;
\draw    (50,104) -- (50,110) ;
\draw    (100,104) -- (100,110) ;
\draw  [dash pattern={on 0.84pt off 2.51pt}]  (50,90) -- (50,104) ;
\draw  [dash pattern={on 0.84pt off 2.51pt}]  (100,90) -- (100,104) ;
\draw  [color={rgb, 255:red, 239; green, 12; blue, 16 }  ,draw opacity=1 ] (87.75,106.28) -- (112.25,106.28) -- (112.25,132.19) -- (87.75,132.19) -- cycle ;
\draw  [draw opacity=0] (10,19.5) -- (140,19.5) -- (140,44.5) -- (10,44.5) -- cycle ;
\draw [color={rgb, 255:red, 239; green, 12; blue, 16 }  ,draw opacity=0.5 ] [dash pattern={on 4.5pt off 4.5pt}]  (112.25,106.28) -- (395,5) ;
\draw [color={rgb, 255:red, 239; green, 12; blue, 16 }  ,draw opacity=0.5 ] [dash pattern={on 4.5pt off 4.5pt}]  (87.75,106.28) -- (220,5) ;
\draw [color={rgb, 255:red, 239; green, 12; blue, 16 }  ,draw opacity=0.5 ] [dash pattern={on 4.5pt off 4.5pt}]  (112.25,132.19) -- (395,220) ;
\draw [color={rgb, 255:red, 239; green, 12; blue, 16 }  ,draw opacity=0.5 ] [dash pattern={on 4.5pt off 4.5pt}]  (87.75,132.19) -- (220,220) ;
\draw  [draw opacity=0] (0,0) -- (400,0) -- (400,225) -- (0,225) -- cycle ;
\draw  [color={rgb, 255:red, 239; green, 12; blue, 16 }  ,draw opacity=1 ][fill={rgb, 255:red, 255; green, 255; blue, 255 }  ,fill opacity=1 ] (220,5) -- (395,5) -- (395,220) -- (220,220) -- cycle ;
\draw    (340,60.2) -- (340,87.2) ;
\draw [shift={(340,90.2)}, rotate = 270] [fill={rgb, 255:red, 0; green, 0; blue, 0 }  ][line width=0.08]  [draw opacity=0] (7.14,-3.43) -- (0,0) -- (7.14,3.43) -- cycle    ;
\draw   (320,110.2) -- (320,115.6) -- (325,116.8) -- (315,119.2) -- (325,121.6) -- (315,124) -- (325,126.4) -- (315,128.8) -- (325,131.2) -- (315,133.6) -- (320,134.8) -- (320,140.2) ;
\draw   (320,95.2) -- (320,79.45) (310,75.95) -- (330,75.95) (320,75.95) -- (320,60.2) (315,80.85) -- (315,79.45) -- (325,79.45) -- (325,80.85) -- (315,80.85) -- cycle ;
\draw  [fill={rgb, 255:red, 0; green, 0; blue, 0 }  ,fill opacity=1 ] (319,183.2) .. controls (319,182.65) and (319.45,182.2) .. (320,182.2) .. controls (320.55,182.2) and (321,182.65) .. (321,183.2) .. controls (321,183.75) and (320.55,184.2) .. (320,184.2) .. controls (319.45,184.2) and (319,183.75) .. (319,183.2) -- cycle ;
\draw  [fill={rgb, 255:red, 0; green, 0; blue, 0 }  ,fill opacity=1 ] (319,147.2) .. controls (319,146.65) and (319.45,146.2) .. (320,146.2) .. controls (320.55,146.2) and (321,146.65) .. (321,147.2) .. controls (321,147.75) and (320.55,148.2) .. (320,148.2) .. controls (319.45,148.2) and (319,147.75) .. (319,147.2) -- cycle ;
\draw    (285,40.2) -- (279.5,40.14) ;
\draw [shift={(276.5,40.11)}, rotate = 0.63] [fill={rgb, 255:red, 0; green, 0; blue, 0 }  ][line width=0.08]  [draw opacity=0] (7.14,-3.43) -- (0,0) -- (7.14,3.43) -- cycle    ;
\draw    (320,95.2) -- (320,110.2) ;
\draw    (340,140.2) -- (340,113.2) ;
\draw [shift={(340,110.2)}, rotate = 90] [fill={rgb, 255:red, 0; green, 0; blue, 0 }  ][line width=0.08]  [draw opacity=0] (7.14,-3.43) -- (0,0) -- (7.14,3.43) -- cycle    ;
\draw    (295,147.2) -- (345,147.2) ;
\draw    (320,140.2) -- (320,147.2) ;
\draw   (295,150.2) -- (295,155.6) -- (300,156.8) -- (290,159.2) -- (300,161.6) -- (290,164) -- (300,166.4) -- (290,168.8) -- (300,171.2) -- (290,173.6) -- (295,174.8) -- (295,180.2) ;
\draw   (345,155.2) -- (345,161.2) (345,175.2) -- (345,169.2) (355,161.2) -- (335,161.2) (355,169.2) -- (335,169.2) ;
\draw    (345,155.2) -- (345,147.2) ;
\draw    (345,183.2) -- (345,175.2) ;
\draw    (295,183.2) -- (345,183.2) ;
\draw    (320,183.2) -- (320,200.2) ;
\draw    (250,200.2) -- (320,200.2) ;
\draw    (250,40.2) -- (320,40.2) ;
\draw    (320,40.2) -- (320,60.2) ;
\draw    (365,180.2) -- (365,153.2) ;
\draw [shift={(365,150.2)}, rotate = 90] [fill={rgb, 255:red, 0; green, 0; blue, 0 }  ][line width=0.08]  [draw opacity=0] (7.14,-3.43) -- (0,0) -- (7.14,3.43) -- cycle    ;
\draw    (250,50.2) -- (250,187.2) ;
\draw [shift={(250,190.2)}, rotate = 270] [fill={rgb, 255:red, 0; green, 0; blue, 0 }  ][line width=0.08]  [draw opacity=0] (7.14,-3.43) -- (0,0) -- (7.14,3.43) -- cycle    ;
\draw  [fill={rgb, 255:red, 255; green, 255; blue, 255 }  ,fill opacity=1 ] (247,40.2) .. controls (247,38.54) and (248.34,37.2) .. (250,37.2) .. controls (251.66,37.2) and (253,38.54) .. (253,40.2) .. controls (253,41.86) and (251.66,43.2) .. (250,43.2) .. controls (248.34,43.2) and (247,41.86) .. (247,40.2) -- cycle ;
\draw  [fill={rgb, 255:red, 255; green, 255; blue, 255 }  ,fill opacity=1 ] (247,200.2) .. controls (247,198.54) and (248.34,197.2) .. (250,197.2) .. controls (251.66,197.2) and (253,198.54) .. (253,200.2) .. controls (253,201.86) and (251.66,203.2) .. (250,203.2) .. controls (248.34,203.2) and (247,201.86) .. (247,200.2) -- cycle ;
\draw    (295,150.2) -- (295,147.2) ;
\draw    (295,183.2) -- (295,180.2) ;
\draw    (167.5,75) -- (167.5,122) ;
\draw [shift={(167.5,125)}, rotate = 270] [fill={rgb, 255:red, 0; green, 0; blue, 0 }  ][line width=0.08]  [draw opacity=0] (7.14,-3.43) -- (0,0) -- (7.14,3.43) -- cycle    ;
\draw    (100,135) -- (140,135) ;
\draw    (164.02,135) -- (140,135) ;
\draw    (85,135) -- (100,135) ;
\draw    (164.02,65) -- (140,65) ;

\draw (169.5,100) node [anchor=west] [inner sep=0.75pt]    {$u_{\text{b}}$};
\draw (61,61.62) node [anchor=south east] [inner sep=0.75pt]  [font=\small]  {$i_{\text{c}}$};
\draw (34,80.01) node [anchor=east] [inner sep=0.75pt]  [font=\small]  {$u_{\text{c}}$};
\draw (152.01,61.6) node [anchor=south] [inner sep=0.75pt]    {$i_{\text{b}}$};
\draw (169.02,138.4) node [anchor=north west][inner sep=0.75pt]    {$-$};
\draw (169.02,53.8) node [anchor=west] [inner sep=0.75pt]    {${\displaystyle +}$};
\draw (75,32) node   [align=left] {Battery};
\draw (342,75.2) node [anchor=west] [inner sep=0.75pt]    {$u_{\text{oc}}$};
\draw (342,125.2) node [anchor=west] [inner sep=0.75pt]    {$u_{\text{Rint}}$};
\draw (367,165.2) node [anchor=west] [inner sep=0.75pt]    {$u_{\text{th}}$};
\draw (248,120.2) node [anchor=east] [inner sep=0.75pt]    {$u_{\text{c}}$};
\draw (313,125.2) node [anchor=east] [inner sep=0.75pt]    {$R_{\text{int}}$};
\draw (288,165.2) node [anchor=east] [inner sep=0.75pt]    {$R_{\text{th}}$};
\draw (333,165.2) node [anchor=east] [inner sep=0.75pt]    {$C_{\text{th}}$};
\draw (283,43.6) node [anchor=north east] [inner sep=0.75pt]    {$i_{\text{c}}$};
\draw (248,36.8) node [anchor=south east] [inner sep=0.75pt]    {$+$};
\draw (318,72.55) node [anchor=south east] [inner sep=0.75pt]    {$+$};
\draw (318,82.85) node [anchor=north east] [inner sep=0.75pt]    {$-$};
\draw (248,203.6) node [anchor=north east] [inner sep=0.75pt]    {$-$};
\draw (222,8) node [anchor=north west][inner sep=0.75pt]   [align=left] {Thevenin model};

\end{tikzpicture}
\caption{Simplified battery circuit \& Thevenin model}
\label{fig:EC_7}
\end{figure}

The first component is the \ac{lib}, whose state is described by the state of charge \(\mathrm{SoC}\) and the battery voltage \(u_\text{b}\). \citet{Hussein(2011)} and \citet{Zhou(2021)} review several modeling approaches for \ac{lib} cells. We adopt an equivalent circuit model due to its descriptive formulation, possible short simulation run-time, and good estimation accuracy. Commonly, \acp{lib} consist of multiple cells, which can be connected in series and parallel, as shown in Fig.\,\ref{fig:EC_7}. For simplification, we assume:

\begin{assumption_new}
\label{ass:Bat1}
All \ac{lib} cells are identical and the load is distributed evenly.
\end{assumption_new}

Considering Assumption \ref{ass:Bat1}, the number of cells connected in series \(N_\text{S}\) and parallel \(N_\text{P}\) define the battery voltage \(u_\text{b}\) and battery current \(i_\text{b}\) by
\begin{equation}
\label{eqn:Ub_Ib}
    u_\text{b} = N_\text{S}\,u_\text{c}, \quad \quad
    i_\text{b} = N_\text{P}\,i_\text{c},
\end{equation}
where \(u_\text{c}\) and \(i_\text{c}\) are the \ac{lib} cells' voltage and current. 

We define the portion of the already discharged battery charge, also called the depth of discharge, as
\begin{equation}
\label{eqn:DoD}
    \mathrm{DoD} =  \mathrm{DoD}_\text{0} + \frac{\eta_\text{b}}{Q_\text{b}} \int i_\text{b}\,\mathrm{d}t,
\end{equation}
where \(\mathrm{DoD}_\text{0}\) is the initial depth of discharge and \(\eta_\text{b}, Q_\text{b}, i_\text{b}\) are the battery's efficiency, capacity and current. This method is called Coulomb counting and is characterized by its simplicity and performance \cite{Nikolian(2014)}. Commonly, this method is used to describe the state of charge 
\begin{equation}
\label{eqn:SoC}
     \mathrm{SoC} = 1 - \mathrm{DoD},
\end{equation}
which is the portion of the remaining battery charge.

For the \ac{lib} cell model, we apply a first-order equivalent circuit model, also known as Thevenin model. It describes the \ac{lib} cell behavior accurately, while the simulation run-time, the complexity, and the needed information about the inner processes of the \ac{lib} cell are limited. Fig.~\ref{fig:EC_7} illustrates the Thevenin model consisting of an ohmic resistance \(R_\text{int}\), an ideal voltage source \(u_\text{oc}\), and an RC parallel network \(R_\text{th}\|C_\text{th}\) connected in series. The total internal resistance is divided into the ohmic resistance \(R_\text{int}\) and the polarization resistance \(R_\text{th}\). If no load is applied, the \ac{lib} cell voltage \(u_\text{c}\) equals the open circuit voltage \(u_\text{oc}\). The polarization RC network describes effects resulting from chemical reactions in the electrode surfaces and the ion mass transfer inside the \ac{lib} cell \citep{He(2011),Nikolian(2014)}. According to Kirchhoff’s circuit laws, we define the characteristic equations of the Thevenin model
\begin{equation*}
\label{eqn:UcUth}
\begin{aligned}
    \frac{\mathrm{d}}{\mathrm{d}t}u_\text{th} &= -\frac{1}{R_\text{th}\,C_\text{th}}\,u_\text{th} + \frac{1}{C_\text{th}}\,i_\text{c},\\
    u_\text{c} &= u_\text{oc} - u_\text{th} - R_\text{int}\,i_\text{c},
\end{aligned}
\end{equation*}
where we insert \eqref{eqn:Ub_Ib}, while considering \eqref{eqn:DoD} to obtain the \ac{lib}'s characteristic equations
\begin{equation}
\label{eqn:Bat}
\begin{aligned}
    \frac{\mathrm{d}}{\mathrm{d}t}\mathrm{DoD} &= \frac{\eta_\text{b}}{Q_\text{b}}  i_\text{b},\\
    \frac{\mathrm{d}}{\mathrm{d}t}u_\text{th} &= -\frac{1}{R_\text{th}\,C_\text{th}}\,u_\text{th} + \frac{1}{N_\text{P}\,C_\text{th}}\,i_\text{b},\\
    u_\text{b} &= N_\text{S} \left(u_\text{oc}  - u_\text{th} - \frac{R_\text{int}}{N_\text{P}}\,i_\text{b}\right).
\end{aligned}
\end{equation}

\begin{remark_new}
Since the polarization effects are only considered by one RC-network, the \ac{lib} cell behavior at the end of discharge phase can not be reproduced accurately.
\end{remark_new}

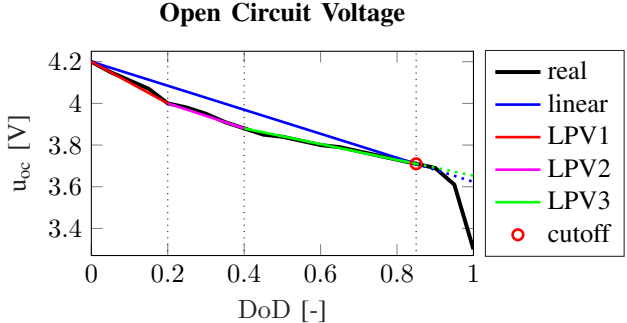
\begin{figure}[H]
\centering
\definecolor{mycolor1}{rgb}{1.00000,0.00000,1.00000}%
\begin{tikzpicture}

\begin{axis}[%
width=2in,
height=1.07in,
at={(0.0in,0.0in)},
scale only axis,
xmin=0,
xmax=1,
xlabel style={font=\color{white!15!black}},
xlabel={\(\mathrm{DoD}\) [-]},
ymin=3.27,
ymax=4.25,
ylabel style={font=\color{white!15!black}},
ylabel={\(\text{u}_{\text{oc}}\) [V]},
axis background/.style={fill=white},
title style={font=\bfseries},
title={Open Circuit Voltage},
legend style={at={(1.03,0.5)}, anchor=west, legend cell align=left, align=left, draw=white!15!black}
]
\addplot [color=black, line width=1.5pt]
  table[row sep=crcr]{%
0	4.2\\
0.05	4.15\\
0.1	4.11\\
0.15	4.07\\
0.2	4\\
0.25	3.98\\
0.3	3.95\\
0.35	3.91\\
0.4	3.88\\
0.45	3.85\\
0.5	3.84\\
0.55	3.82\\
0.6	3.8\\
0.65	3.79\\
0.7	3.77\\
0.75	3.75\\
0.8	3.73\\
0.85	3.71\\
0.9	3.69\\
0.95	3.61\\
1	3.3\\
};
\addlegendentry{real}

\addplot [color=blue, line width=1.0pt]
  table[row sep=crcr]{%
0	4.2\\
0.85	3.71\\
};
\addlegendentry{linear}

\addplot [color=red, line width=1.0pt]
  table[row sep=crcr]{%
0	4.2\\
0.2	4\\
};
\addlegendentry{LPV1}

\addplot [color=mycolor1, line width=1.0pt]
  table[row sep=crcr]{%
0.2	4\\
0.4	3.88\\
};
\addlegendentry{LPV2}

\addplot [color=green, line width=1.0pt]
  table[row sep=crcr]{%
0.4	3.88\\
0.85	3.71\\
};
\addlegendentry{LPV3}

\addplot [color=red, line width=1.0pt, only marks, mark=o, mark options={solid, red}]
  table[row sep=crcr]{%
0.85	3.71\\
};
\addlegendentry{cutoff}

\addplot [color=blue, dotted, line width=1.0pt, forget plot]
  table[row sep=crcr]{%
0.85	3.71\\
1	3.62352941176471\\
};
\addplot [color=green, dotted, line width=1.0pt, forget plot]
  table[row sep=crcr]{%
0.85	3.71\\
1	3.65333333333333\\
};
\addplot [color=black, dotted, forget plot]
  table[row sep=crcr]{%
0.2	3.2\\
0.2	4.25\\
};
\addplot [color=black, dotted, forget plot]
  table[row sep=crcr]{%
0.4	3.2\\
0.4	4.25\\
};
\addplot [color=black, dotted, forget plot]
  table[row sep=crcr]{%
0.85	3.2\\
0.85	4.25\\
};
\end{axis}
\end{tikzpicture}%
\caption{Generalized discharge curve of a LiPo cell (black) \citep{Florea(2020)}, linear approximation (blue), and \ac{lpv} approximations (red, magenta, green)}
\vspace{-0.25cm}
\label{fig:OCV}
\end{figure}

To increase the accuracy of the \ac{lib} cell model, we define the ideal voltage source \(u_\text{oc}\) as dependent on the depth of discharge \(\mathrm{DoD}\). Fig.~\ref{fig:OCV} shows a generalized discharge curve of a \ac{lib} cell (black graph) and the cutoff depth of discharge \(\overline{\mathrm{DoD}}_\text{cutoff}\) (red dot), set to \(85\%\), which limits the depth of discharge to avoid over-discharging. 

\begin{remark_new}
The actual discharge curve of a \ac{lib} cell is highly individual and can differ from the black graph in Fig.~\ref{fig:OCV} due to external circumstances (temperature, state of health, discharge rate) and technological differences, et cetera. 
\end{remark_new}

According to the blue graph, the linear approximation
\begin{equation}
\label{eqn:b0b1_lin}
    u_\text{oc} = b_\text{0} + b_\text{1}\,\mathrm{DoD} \ \ \ \text{for} \ \ \ 0 \leq \mathrm{DoD} \leq \overline{\mathrm{DoD}}_\text{cutoff}.
\end{equation}
is parameterized by the open circuit voltage parameters \(b_\text{0}\) and \(b_\text{1}\).
Due to the non-linearity of the discharge curve, we propose to define a piece-wise linear function
\begin{equation}
\label{eqn:b0b1_LPV}
u_\text{oc} = \Bigg\{
\begin{array}{l}
b_\text{0,1} + b_\text{1,1}\,\mathrm{DoD} \ \text{for} \ \overline{\mathrm{DoD}}_0 \leq \mathrm{DoD} \leq \overline{\mathrm{DoD}}_1, \\
b_\text{0,2} + b_\text{1,2}\,\mathrm{DoD} \ \text{for} \ \overline{\mathrm{DoD}}_1 \leq \mathrm{DoD} \leq \overline{\mathrm{DoD}}_2, \\
b_\text{0,3} + b_\text{1,3}\,\mathrm{DoD} \ \text{for} \ \overline{\mathrm{DoD}}_2 \leq \mathrm{DoD} \leq \overline{\mathrm{DoD}}_3,
\end{array}
\end{equation}
where the index \(i \in \{1,\dots,3\}\) indicates the active \ac{lpv} battery model. Depending on this index the corresponding parameters \(b_{\text{0},i}\), \(b_{\text{1},i}\) and the thresholds \(\left[\,\overline{\mathrm{DoD}}_{i-1},\overline{\mathrm{DoD}}_i\right]\) are chosen to fit the red-, magenta-, and green-colored graphs in Fig.~\ref{fig:OCV}, respectively.

\subsection{Electric Speed Controller}
\label{sec:ESC}
The second component is the \ac{esc}, which connects the \ac{lib} with the \ac{bldc} motor and controls the \ac{bldc} motor, depending on the \ac{pwm} control command \(s_\text{PWM}\). Since the \ac{bldc} motor will be modeled as a simplified \ac{dc} motor, we model the \ac{esc} as a DC-to-DC converter, which regulates the voltage supply to a \ac{dc} motor. 
Then, the \ac{esc} converts the battery voltage \(u_\text{b}\) to the \ac{dc} motor's voltage 
\begin{equation*}
\label{eqn:u_esc}
u_\text{DC} = f_\text{ESC}(s_\text{PWM})\,u_\text{b},
\end{equation*} 
depending on the function \(f_\text{ESC}(s_\text{PWM})\), which is approximately a linear function \citep{Michel(2019)}.
In order to formulate the relation between the \ac{dc} motor's supply power and the \ac{esc}'s input current 
\begin{equation}
\label{eqn:IESCb}
i_\text{ESC} = \frac{1}{\eta_\text{ESC}}f_\text{ESC}(s_\text{PWM})\,i_\text{DC} = \frac{p_\text{DC}}{\eta_\text{ESC}\,u_\text{b}},
\end{equation} 
 we consider the \ac{esc}'s efficiency \(\eta_\text{ESC}\) and follow the energy conservation law
\begin{equation*}
p_\text{ESC}\,\eta_\text{ESC} = p_\text{DC},
\end{equation*}
where \(p_\text{ESC}=i_\text{ESC}\,u_\text{b}\) is the power supplied to the \ac{esc}.

\subsection{Brushless Direct Current Motor with a fixed Rotor}
\label{sec:BLDC}
The third component is the \ac{bldc} motor. It is a special kind of synchronous machine, which is controlled by an \ac{esc}. According to a commutation logic, which depends on the rotor position, direct currents are applied to the three input wires of the \ac{bldc} motor. By this, the magnetic poles of the stator coils align with the rotor monopoles to initiate or maintain the rotation of the rotor. For more information, see \citep{Babiel(2020)}. 
\vspace{-0.25cm}
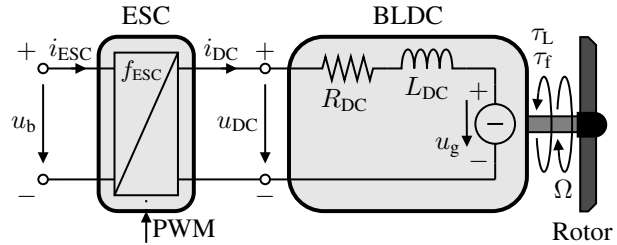
\begin{figure}[H]
\centering
\tikzset{every picture/.style={line width=0.75pt}} %

\begin{tikzpicture}[x=0.6pt,y=0.6pt,yscale=-1,xscale=1]

\draw  [fill={rgb, 255:red, 128; green, 128; blue, 128 }  ,fill opacity=1 ] (330,80.2) -- (363.6,80.2) -- (363.6,90.2) -- (330,90.2) -- cycle ;
\draw  [fill={rgb, 255:red, 155; green, 155; blue, 155 }  ,fill opacity=0.25 ][line width=1.5]  (180,52) .. controls (180,39.85) and (189.85,30) .. (202,30) -- (308,30) .. controls (320.15,30) and (330,39.85) .. (330,52) -- (330,118) .. controls (330,130.15) and (320.15,140) .. (308,140) -- (202,140) .. controls (189.85,140) and (180,130.15) .. (180,118) -- cycle ;
\draw  [fill={rgb, 255:red, 74; green, 74; blue, 74 }  ,fill opacity=1 ] (363.6,140.2) -- (363.6,90.2) -- (368.6,85.2) -- (373.6,85.2) -- (373.6,85.2) -- (373.6,135.2) -- (368.6,140.2) -- (363.6,140.2) -- cycle ;
\draw  [fill={rgb, 255:red, 74; green, 74; blue, 74 }  ,fill opacity=1 ] (373.6,35.2) -- (373.6,85.2) -- (373.6,85.2) -- (368.6,85.2) -- (363.6,80.2) -- (363.6,30.2) -- (363.6,30.2) -- (368.6,30.2) -- cycle ;
\draw  [fill={rgb, 255:red, 0; green, 0; blue, 0 }  ,fill opacity=1 ] (372.5,78) .. controls (376.64,78) and (380,81.36) .. (380,85.5) -- (380,85.5) .. controls (380,89.64) and (376.64,93) .. (372.5,93) -- (362,93) .. controls (362,93) and (362,93) .. (362,93) -- (362,78) .. controls (362,78) and (362,78) .. (362,78) -- cycle ;
\draw    (181,120) -- (165,120) ;
\draw    (180,50) -- (192,50) ;
\draw   (192,50) -- (201,50) -- (203,42.5) -- (207,57.5) -- (211,42.5) -- (215,57.5) -- (219,42.5) -- (223,57.5) -- (227,42.5) -- (231,57.5) -- (233,50) -- (242,50) ;
\draw   (242,50) -- (251,50) .. controls (251,43.38) and (252.79,38) .. (255,38) .. controls (257.21,38) and (259,43.38) .. (259,50) .. controls (259,43.38) and (260.79,38) .. (263,38) .. controls (265.21,38) and (267,43.38) .. (267,50) .. controls (267,43.38) and (268.79,38) .. (271,38) .. controls (273.21,38) and (275,43.38) .. (275,50) .. controls (275,43.38) and (276.79,38) .. (279,38) .. controls (281.21,38) and (283,43.38) .. (283,50) -- (292,50) ;
\draw   (310,97.5) .. controls (303.1,97.5) and (297.5,91.9) .. (297.5,85) .. controls (297.5,78.1) and (303.1,72.5) .. (310,72.5) .. controls (316.9,72.5) and (322.5,78.1) .. (322.5,85) .. controls (322.5,91.9) and (316.9,97.5) .. (310,97.5) -- cycle (310,110) -- (310,97.5) (310,60) -- (310,72.5) ;
\draw    (292,50) -- (310,50) ;
\draw    (310,50) -- (310,60) ;
\draw    (180,120) -- (310,120) ;
\draw    (310,110) -- (310,120) ;
\draw [line width=0.75]    (137.5,50) -- (144.5,50) ;
\draw [shift={(147.5,50)}, rotate = 180] [fill={rgb, 255:red, 0; green, 0; blue, 0 }  ][line width=0.08]  [draw opacity=0] (7.14,-3.43) -- (0,0) -- (7.14,3.43) -- cycle    ;
\draw [line width=0.75]    (292,69.92) -- (292.15,96.56) ;
\draw [shift={(292.17,99.56)}, rotate = 269.68] [fill={rgb, 255:red, 0; green, 0; blue, 0 }  ][line width=0.08]  [draw opacity=0] (7.14,-3.43) -- (0,0) -- (7.14,3.43) -- cycle    ;
\draw    (165,60) -- (165,107) ;
\draw [shift={(165,110)}, rotate = 270] [fill={rgb, 255:red, 0; green, 0; blue, 0 }  ][line width=0.08]  [draw opacity=0] (7.14,-3.43) -- (0,0) -- (7.14,3.43) -- cycle    ;
\draw  [draw opacity=0] (337.22,65.1) .. controls (338.09,59.02) and (339.35,55.2) .. (340.75,55.2) .. controls (343.37,55.2) and (345.5,68.63) .. (345.5,85.2) .. controls (345.5,101.77) and (343.37,115.2) .. (340.75,115.2) .. controls (338.61,115.2) and (336.8,106.25) .. (336.2,93.94) -- (340.75,85.2) -- cycle ; \draw   (337.22,65.1) .. controls (338.09,59.02) and (339.35,55.2) .. (340.75,55.2) .. controls (343.37,55.2) and (345.5,68.63) .. (345.5,85.2) .. controls (345.5,101.77) and (343.37,115.2) .. (340.75,115.2) .. controls (338.61,115.2) and (336.8,106.25) .. (336.2,93.94) ;  
\draw    (337.22,65.1) -- (336.32,73.5) ;
\draw [shift={(336,76.48)}, rotate = 276.14] [fill={rgb, 255:red, 0; green, 0; blue, 0 }  ][line width=0.08]  [draw opacity=0] (7.14,-3.43) -- (0,0) -- (7.14,3.43) -- cycle    ;
\draw  [draw opacity=0] (349.24,75.83) .. controls (349.86,63.85) and (351.64,55.2) .. (353.75,55.2) .. controls (356.37,55.2) and (358.5,68.63) .. (358.5,85.2) .. controls (358.5,101.77) and (356.37,115.2) .. (353.75,115.2) .. controls (352.43,115.2) and (351.24,111.82) .. (350.38,106.36) -- (353.75,85.2) -- cycle ; \draw   (349.24,75.83) .. controls (349.86,63.85) and (351.64,55.2) .. (353.75,55.2) .. controls (356.37,55.2) and (358.5,68.63) .. (358.5,85.2) .. controls (358.5,101.77) and (356.37,115.2) .. (353.75,115.2) .. controls (352.43,115.2) and (351.24,111.82) .. (350.38,106.36) ;  
\draw    (350.38,106.36) -- (349.63,97.35) ;
\draw [shift={(349.38,94.36)}, rotate = 85.24] [fill={rgb, 255:red, 0; green, 0; blue, 0 }  ][line width=0.08]  [draw opacity=0] (7.14,-3.43) -- (0,0) -- (7.14,3.43) -- cycle    ;
\draw    (315,85) -- (305,85.02) ;
\draw  [fill={rgb, 255:red, 155; green, 155; blue, 155 }  ,fill opacity=0.25 ][line width=1.5]  (60,41.99) .. controls (60,35.36) and (65.37,29.99) .. (72,29.99) -- (108,29.99) .. controls (114.63,29.99) and (120,35.36) .. (120,41.99) -- (120,127.99) .. controls (120,134.61) and (114.63,139.99) .. (108,139.99) -- (72,139.99) .. controls (65.37,139.99) and (60,134.61) .. (60,127.99) -- cycle ;
\draw    (25,50) -- (70,50) ;
\draw    (42.5,49.99) -- (49.5,50) ;
\draw [shift={(52.5,50)}, rotate = 180.08] [fill={rgb, 255:red, 0; green, 0; blue, 0 }  ][line width=0.08]  [draw opacity=0] (7.14,-3.43) -- (0,0) -- (7.14,3.43) -- cycle    ;
\draw   (70,39.99) -- (110,39.99) -- (110,129.99) -- (70,129.99) -- cycle ;
\draw    (25,120.01) -- (70,120) ;
\draw    (110,39.99) -- (70,129.99) ;
\draw    (25,60) -- (25,107) ;
\draw [shift={(25,110)}, rotate = 270] [fill={rgb, 255:red, 0; green, 0; blue, 0 }  ][line width=0.08]  [draw opacity=0] (7.14,-3.43) -- (0,0) -- (7.14,3.43) -- cycle    ;
\draw    (165,120) -- (110,120) ;
\draw    (181,50) -- (165,50) ;
\draw    (165,50) -- (110,50) ;
\draw  [fill={rgb, 255:red, 255; green, 255; blue, 255 }  ,fill opacity=1 ] (168,50) .. controls (168,51.66) and (166.66,53) .. (165,53) .. controls (163.34,53) and (162,51.66) .. (162,50) .. controls (162,48.34) and (163.34,47) .. (165,47) .. controls (166.66,47) and (168,48.34) .. (168,50) -- cycle ;
\draw  [fill={rgb, 255:red, 255; green, 255; blue, 255 }  ,fill opacity=1 ] (168,120) .. controls (168,121.66) and (166.66,123) .. (165,123) .. controls (163.34,123) and (162,121.66) .. (162,120) .. controls (162,118.34) and (163.34,117) .. (165,117) .. controls (166.66,117) and (168,118.34) .. (168,120) -- cycle ;
\draw  [fill={rgb, 255:red, 255; green, 255; blue, 255 }  ,fill opacity=1 ] (28,120.01) .. controls (28,121.67) and (26.66,123.01) .. (25,123.01) .. controls (23.34,123.01) and (22,121.67) .. (22,120.01) .. controls (22,118.36) and (23.34,117.01) .. (25,117.01) .. controls (26.66,117.01) and (28,118.36) .. (28,120.01) -- cycle ;
\draw  [fill={rgb, 255:red, 255; green, 255; blue, 255 }  ,fill opacity=1 ] (28,50) .. controls (28,51.66) and (26.66,53) .. (25,53) .. controls (23.34,53) and (22,51.66) .. (22,50) .. controls (22,48.34) and (23.34,47) .. (25,47) .. controls (26.66,47) and (28,48.34) .. (28,50) -- cycle ;
\draw  [draw opacity=0] (60,4) -- (120,4) -- (120,29) -- (60,29) -- cycle ;
\draw  [draw opacity=0] (180,3) -- (330,3) -- (330,28) -- (180,28) -- cycle ;
\draw  [draw opacity=0] (330,140) -- (400,140) -- (400,165) -- (330,165) -- cycle ;
\draw  [draw opacity=0] (0,0) -- (400,0) -- (400,170) -- (0,170) -- cycle ;
\draw    (90,160) -- (90,143.03) ;
\draw [shift={(90,140.03)}, rotate = 90] [fill={rgb, 255:red, 0; green, 0; blue, 0 }  ][line width=0.08]  [draw opacity=0] (7.14,-3.43) -- (0,0) -- (7.14,3.43) -- cycle    ;
\draw  [dash pattern={on 0.84pt off 2.51pt}]  (90,140.03) -- (90,130.03) ;

\draw (165,123.4) node [anchor=north] [inner sep=0.75pt]    {$-$};
\draw (165,46.6) node [anchor=south] [inner sep=0.75pt]    {${\displaystyle +}$};
\draw (290.17,99.56) node [anchor=east] [inner sep=0.75pt]    {$u_{\text{g}}$};
\draw (353,118.4) node [anchor=north] [inner sep=0.75pt]    {$\Omega $};
\draw (340,51.6) node [anchor=south] [inner sep=0.75pt]    {$\tau _{\text{f}}$};
\draw (137.5,46.6) node [anchor=south] [inner sep=0.75pt]  [font=\normalsize]  {$i_{\text{DC}}$};
\draw (215,60.9) node [anchor=north] [inner sep=0.75pt]    {$R_{\text{DC}}$};
\draw (267,53.4) node [anchor=north] [inner sep=0.75pt]    {$L_{\text{DC}}$};
\draw (308,71.6) node [anchor=south east] [inner sep=0.75pt]    {${\displaystyle +}$};
\draw (308,100.9) node [anchor=north east] [inner sep=0.75pt]    {$-$};
\draw (23,85) node [anchor=east] [inner sep=0.75pt]    {$u_{\text{b}}$};
\draw (42.5,46.59) node [anchor=south] [inner sep=0.75pt]  [font=\normalsize]  {$i_{\text{ESC}}$};
\draw (163,85) node [anchor=east] [inner sep=0.75pt]    {$u_{\text{DC}}$};
\draw (72,43.39) node [anchor=north west][inner sep=0.75pt]  [font=\small]  {$f_{\text{ESC}}$};
\draw (23,123.41) node [anchor=north east] [inner sep=0.75pt]    {$-$};
\draw (23,46.6) node [anchor=south east] [inner sep=0.75pt]    {${\displaystyle +}$};
\draw (340,36.6) node [anchor=south] [inner sep=0.75pt]    {$\tau _{\text{L}}$};
\draw (90,16.5) node   [align=left] {\acs{esc}};
\draw (255,16.5) node   [align=left] {\acs{bldc}};
\draw (365,152.5) node   [align=left] {Rotor};
\draw (92,143.43) node [anchor=north west][inner sep=0.75pt]    {PWM};

\end{tikzpicture}
\vspace{-0.25cm}
\caption{Simplified \acs{esc-bldc} circuit}
\label{fig:EC_6}
\end{figure}
\vspace{-0.25cm}
Fig.~\ref{fig:EC_6} shows a simplified \acs{esc-bldc} circuit. Since the structure and dynamics of a \ac{bldc} motor are complex and not continuous, we approximate its power consumption by a simplified \ac{dc} motor with a fixed rotor, based on \citep{Fouad(2017),Li(2022),Morbidi(2016)}. Here, we assume: 
\begin{assumption_new}
\label{ass:BLDC_1}
The \ac{bldc} motor is driving at a constant speed and the motor friction torque is negligible due to liquid lubrication.
\end{assumption_new}
As shown in Fig.~\ref{fig:EC_6}, the simplified \ac{dc} motor is built from elementary electrical components. The resistances of the motor and the inductance of the coils are summarized as the motor's internal resistance \(R_\text{DC}\) and the motor's inductance \(L_\text{DC}\). An ideal power sink \(u_\text{g} = \Omega/K_\text{V}\) represents the transformation of electrical power to mechanical power, where \(K_\text{V}\) is the voltage constant of the motor and \(\Omega\) is the motor speed. According to Kirchhoff's circuit laws, the motor voltage \(u_\text{DC}\) is given by 
\begin{equation}
\label{eqn:BLDC_U1}
\begin{split}
    u_\text{DC} = u_\text{R} + u_\text{L} + u_\text{g} &= R_\text{DC}\,i_\text{DC} + L_\text{DC}\frac{\mathrm{d}\,i_\text{DC}}{\mathrm{d}\,t} + \frac{1}{K_\text{V}} \Omega \\ &= R_\text{DC}\,i_\text{DC} + \frac{1}{K_\text{V}} \Omega,
\end{split}
\end{equation}
where the motor current reads
\begin{equation}
\label{eqn:BLDC_I1}
\begin{split}
    i_\text{DC} = \frac{1}{K_\tau}\tau_M &= \frac{1}{K_\tau}\left(J_\text{r}\frac{\mathrm{d}\,\Omega(t)}{\mathrm{d}t} + D_\text{f}\,\Omega + \tau_\text{f} + \tau_\text{L}\right) \\ &= K_\text{V}\,\left( \,D_\text{f}\,\Omega + \tau_\text{L}\right).
\end{split}
\end{equation}
It depends on the motor torque \(\tau_\text{M}\) and the motor's torque constant \(K_\tau\), which is approximated by \(K_\tau = 1/K_\text{V}\). The motor torque \(\tau_\text{M}\) represents the torque required to change the motor speed \(J_\text{r}\frac{\mathrm{d}\,\Omega}{\mathrm{d}t}\) and to compensate for the viscous damping of the motor \(D_\text{f}\,\Omega\), the motor friction torque \(\tau_\text{f}\) and the load friction torque \(\tau_\text{L}\). Here, \(J_\text{r}\) is the moment of inertia of the rotor and \(D_\text{f}\) is the viscous damping factor of the motor. According to \eqref{eqn:Mi}, the load friction torque \(\tau_\text{L} = k_\text{M}\,\Omega^2\) equals the aerodynamic torque produced by the rotor. Equations \eqref{eqn:BLDC_U1} and \eqref{eqn:BLDC_I1} are simplified according to Assumption \ref{ass:BLDC_1} to obtain the power consumption of the \ac{bldc} motor 
\begin{equation}
\label{eqn:BLDC_P1}
\begin{aligned}
p_\text{DC} &= u_\text{DC}\,i_\text{DC} \\
&= \underbrace{R_\text{DC}\,K_\text{V}^2\,\left(D_\text{f}\,\Omega + k_\text{M}\,\Omega^2\right)^2}_{P_\text{el}} + \underbrace{D_\text{f}\,\Omega^2}_{P_\text{mech}} + \underbrace{k_\text{M}\,\Omega^3}_{P_\text{out}}.        
\end{aligned}
\end{equation}
It consists of the electrical power loss \(P_\text{el}\), the mechanical power loss \(P_\text{mech}\) and the mechanical output power \(P_\text{out}\).

\subsection{Combined Energy Consumption Model}
\label{sec:ModelECM}
We combine all three components of the power train to derive the \ac{ecm} for electric-propelled \acp{uav}. Fig.~\ref{fig:EC_0} illustrates the standard circuit, where the \acs{esc-bldc} connections are connected in parallel to the battery. To derive the \ac{ecm}, we define the motor speeds and external power consumption \(\mathbf{u}_\text{e} = (\Omega_1, \dots, \Omega_{N_\text{M}})^\top\) as input and the state contains the depth of discharge \(\mathrm{DoD}\), and the polarization voltage \(u_\text{th}\). Further, the output contains the state of charge \(\mathrm{SoC}\), the battery voltage \(u_\text{b}\) and the battery current \(i_\text{b}\). 

Considering \eqref{eqn:Bat}, \eqref{eqn:IESCb} and \eqref{eqn:BLDC_P1}, the battery current 
\begin{equation*}
\label{eqn:ib_0}
i_\text{b} 
= \sum_{i=1}^{N_\text{M}} i_{\text{ESC},i} 
= \frac{1}{u_\text{b}(\mathrm{DoD}, u_\text{th}, i_\text{b})}\,\sum_{i=1}^{N_\text{M}} \frac{p_{\text{DC},i}(\Omega_i)}{\eta_{\text{ESC},i}},
\end{equation*}
is defined as the sum of the input currents of the \acp{esc} \(i_{\text{ESC},i}\), where \(N_\text{M}\) defines the number of \ac{bldc} motors. If the \ac{bldc} motors do not consume any power, the battery current equals zero. Therefore, we get
\begin{equation}
\label{eqn:ib}
i_\text{b} = \tilde{i} - \sqrt{\tilde{i}^2-\frac{N_\text{P}}{N_\text{S}\,R_\text{int}}\,\sum_{i=1}^{N_\text{M}} \frac{p_{\text{DC},i}(\Omega_i)}{\eta_{\text{ESC},i}}},
\end{equation}
where \(\tilde{i}\) substitutes for \(\tilde{i} = \frac{N_\text{P}\,(u_\text{oc}(\mathrm{DoD}) - u_\text{th})}{2\,R_\text{int}}\).
\begin{remark_new}
The power consumption of other components, like the avionics, the payload, or additional actuators can be added in \eqref{eqn:ib} to the sum of power terms if they consume a significant amount of power, compared to the \ac{bldc} motors.
\end{remark_new}

With \eqref{eqn:DoD}, \eqref{eqn:SoC}, \eqref{eqn:Bat} and \eqref{eqn:ib}, we obtain the nonlinear state space \ac{ecm} for electric propelled \acp{uav} by
\begin{equation}
\label{eqn:Nonlin_ECM}
\begin{split}
    \dot{\mathbf{x}}_\text{e} &= \mathbf{f}_\text{e}(\mathbf{x}_\text{e},\mathbf{u}_\text{e}) + \boldsymbol{\Gamma}_\text{e,x}, \\
    \mathbf{y}_\text{e} &= \mathbf{g}_\text{e}(\mathbf{x}_\text{e},\mathbf{u}_\text{e}) + \boldsymbol{\Gamma}_\text{e,y},
\end{split}
\end{equation}
with the state \(\mathbf{x}_\text{e} = (\mathrm{DoD}, u_\text{th})^\top\), the input \(\mathbf{u}_\text{e} = (\Omega_1,\dots,\Omega_{N_\text{M}})^\top\) and the output \(\mathbf{y}_\text{e} = (\mathrm{SoC}, u_\text{b}, i_\text{b})^\top\), where the right-hand sides of \eqref{eqn:Nonlin_ECM} read 
\begin{align*}
    \mathbf{f}_\text{e}(\mathbf{x}_\text{e},\mathbf{u}_\text{e}) &= 
    \begin{pmatrix} 
    \cfrac{\eta_\text{b}}{Q_\text{b}}\, i_\text{b}(\cdot) \\ 
    -\cfrac{1}{R_\text{th}\,C_\text{th}}\,u_\text{th} + \cfrac{1}{N_\text{P}\,C_\text{th}}\,i_\text{b}(\cdot)
    \end{pmatrix}, %
    \\
    \mathbf{g}_\text{e}(\mathbf{x}_\text{e},\mathbf{u}_\text{e}) &= 
    \begin{pmatrix} 
    1-\mathrm{DoD} \\ 
    N_\text{S}\left(u_\text{oc}(\mathrm{DoD}) - u_\text{th} - \cfrac{R_\text{int}}{N_\text{P}}\,i_\text{b}(\cdot)\right) \\
    i_\text{b}(\cdot) 
    \end{pmatrix}. \nonumber %
\end{align*}
Here, \(i_\text{b}(\mathrm{DoD}, u_\text{th},\Omega_1, \dots, \Omega_{N_\text{M}})\) is defined by \eqref{eqn:ib} and \(u_\text{oc}(\mathrm{DoD})\) can be ether selected according to \eqref{eqn:b0b1_lin} or \eqref{eqn:b0b1_LPV} for the nonlinear or \ac{npv} \ac{ecm}. Further, \(\boldsymbol{\Gamma}_\text{e,x}\) and \(\boldsymbol{\Gamma}_\text{e,y}\) represent uncertainties, originating from modeling inaccuracies and external factors, as temperature changes.

Considering Assumption \ref{ass:UAV_2}, the general \ac{ecm} for electric-propelled \acp{uav} is adjusted especially for multicopters to derive a fitting discrete-time \ac{lpv} \ac{ecm} in Appendix \ref{cha:AppendixD}. For the chosen set point, the hovering state with a fully charged battery \(\mathbf{x}_\text{e}=(0,\dots,0)^\top\), the thrust \(T_\text{SP}\) compensates for the weight force, resulting in equal motor seeds \(\Omega_{\text{SP}} = \sqrt{\frac{m\,\text{g}}{k_\text{F}\,N_\text{M}}}\) for all \(N_\text{M}\) motors. Due to \eqref{eqn:T} and the properties of linear models, the input \(\mathbf{u}_\text{e}\) is reduced to \[\tilde{\mathbf{u}}_\text{e} = \Delta T = T - T_\text{SP},\] which is the deviation of the thrust from the set point.  After the linearization, discretization, and reduction, we obtain the discrete-time linear state-space multicopter \ac{ecm} 
\begin{equation*}
\label{eqn:Lin_ECM}
\begin{aligned}
    \mathbf{x}_\text{e}(k+1) &= \mathbf{A}_\text{d,e}\,\mathbf{x}_\text{e}(k)+\mathbf{B}_\text{d,e}\,\tilde{\mathbf{u}}_\text{e}(k)+\mathbf{E}_\text{d,e} + \boldsymbol{\Gamma}_\text{e,x}, \\
    \mathbf{y}_\text{e}(k) &= \mathbf{C}_\text{d,e}\,\mathbf{x}_\text{e}(k)+\mathbf{D}_\text{d,e}\,\tilde{\mathbf{u}}_\text{e}(k) + \mathbf{y}_\text{e,SP} + \boldsymbol{\Gamma}_\text{e,y}, 
    \end{aligned}
\end{equation*}
where \(\mathbf{A}_\text{d,e}\), \(\mathbf{B}_\text{d,e}\), \(\mathbf{C}_\text{d,e}\) and \(\mathbf{D}_\text{d,e}\) are the discrete-time state-space matrices. Since the set point is not an equilibrium point, \(\mathbf{E}_\text{d,e} = \mathbf{f}_\text{e}(\mathbf{x}_\text{e,SP}, \mathbf{u}_\text{c,SP})\,\Delta t\) and \(\mathbf{y}_\text{e,SP} = \mathbf{g}_\text{e}(\mathbf{x}_\text{e,SP}, \mathbf{u}_\text{c,SP})\) are added as an offset for the energy consumption during hovering. 

\subsection{Power Train Capabilities}
\label{sec:ECM_constr}
In order to prevent damage to the battery, the output of the \ac{ecm} should be constrained by
\begin{alignat*}{2}
\label{constr:ECM}
\mathrm{SoC}_\text{cutoff} &\leq \mathrm{SoC} &&\leq 1,\nonumber\\
u_\text{min}\,N_\text{S} &\leq \ u_\text{b} &&\leq u_\text{max}\,N_\text{S},\\
-i_\text{charge,max} &\leq \ i_\text{b} &&\leq i_\text{discharge,max}, \nonumber
\end{alignat*}
where \(\mathrm{SoC}_\text{cutoff}\) is the cutoff state of charge, \((u_\text{min}, u_\text{max})\) are the voltage boundaries for a \ac{lib} cell and \((i_\text{charge,max}, i_\text{discharge,max})\) are the upper bound of the currents during charge and discharge.

\section{Sensor and Communication Models}
\label{cha:sensor}
Depending on the mission, \acp{uav} are equipped with additional sensors, which often impose additional constraints on the distance to a given target to capture or the velocity of the \ac{uav}. For example, during surveillance missions it is necessary to maintain a maximum distance to the ground and a maximum ground velocity to ensure that the measurements are valid and complete. The most common sensor types are cameras and \acp{lidar}. Cameras are widely used for visual tasks such as aerial imaging and object detection, offering high resolution but are sensitive to lighting and weather conditions. \acp{lidar}, in contrast, provide precise 3D mapping and perform well in low-visibility environments, though they are more data-intensive and costly. Each sensor imposes requirements on the \ac{uav} operations and the mission planning. Therefore, we derive the corresponding constraints for both sensor types in the following.

\subsection{Camera Model}
\label{sec:camera}
When using a camera for data collection, both the alignment with the target and the spatial resolution of the image \(R_\text{I}\), must be considered. The spatial resolution for an image taken from a distance \(d\)
\begin{equation*}
\label{eqn:R_I}
R_\text{I} = I/L = \frac{I}{2\,d\,\tan\left(\gamma/2\right)},
\end{equation*}
which is expressed in pixels per meter, depends on the camera’s image resolution \(I\) and field of view \(\gamma\). To ensure a minimum spatial resolution \(R_\text{I,min}\), the \ac{uav}'s distance \(d_\text{t}\) to its target must satisfy
\begin{align}
\label{eqn:constr_dt}
d_\text{t} = \|\mathbf{p}_\text{t}-\mathbf{p}\|_2 \leq \frac{I}{2\,R_\text{I,min}\,\tan\left(\gamma/2\right)},
\end{align}
where \(\mathbf{p}\) and \(\mathbf{p}_\text{t}\) are the positions of the \ac{uav} and its target.

Further, the alignment of the camera is defined by the angle
\begin{align*}
\chi = \arccos\left(\frac{\mathbf{p}_\text{t}-\mathbf{p}}{\|\mathbf{p}_\text{t}-\mathbf{p}\|_2}\,\mathbf{R}_\text{B}^\text{I}\,\mathbf{a}_\text{c}^\text{B}\right),
\end{align*}
which measures how well the camera’s view aligns with the target. Here, \(\mathbf{a}_\text{c}^\text{B}\) is the normalized camera mounting vector in the body-fixed frame and the target gets centered in the image, when the alignment angle \(\chi\) is minimized.

\begin{figure}[H]
\centering
\input{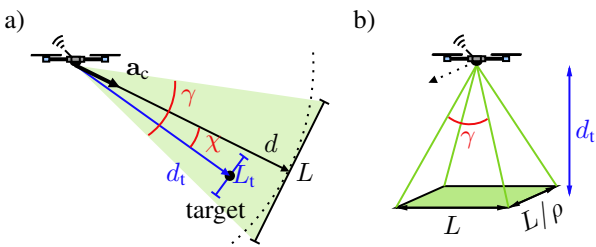}
\vspace{-0.25cm}
\caption{a) Fixed camera; b) Controlled camera facing towards ground;}
\label{fig:Camera}
\end{figure}

For a fixed camera setup (see Fig.~\ref{fig:Camera}a), capturing the target and the surrounding area of interest requires
that the alignment angle remains below a threshold defined by 
\begin{align*}
\chi \leq \gamma/2 - \arctan\left(\frac{L_\text{t}}{2\,d_\text{t}}\right),
\end{align*}
where \(L_\text{t} \leq I/R_\text{I,min}\) is the diameter of the area of interest.

In a simpler case, the camera is mounted on a controlled gimbal, adjusting \(\mathbf{a}_\text{c}^\text{B}\) to track the target. In this case \(\chi \approx 0\) and it is sufficient to maintain the \ac{uav}'s flight altitude \(z \leq z_\text{t}\) above the target altitude \(z_\text{t}\). 

During surveillance missions, such as described by \citet{DiFranco(2015)} and shown in Fig.~\ref{fig:Camera}b, the camera is often controlled to point towards the ground with the target distance \(d_\text{t} = |z_\text{t}-z|\). Increasing the \ac{uav}'s flight altitude, while considering \eqref{eqn:constr_dt}, allows for a larger captured area. Depending on the image's aspect ratio \(\rho\), the \ac{uav}'s ground velocity \(v_\text{g}\) is constrained by
\begin{align*}
v_\text{g} \leq \frac{2\,|z_\text{t}-z|\,\tan\left(\gamma/2\right)\,(1-\delta_\text{c})}{\rho\,T_\text{s,c}},
\end{align*}
where \(\delta_\text{c} \in [0,1]\) is the overlap of successive images and \(T_\text{s,c}\) is camera's sampling period. 
\subsection{LIDAR Model}
\label{sec:lidar}
\begin{figure}[H]
\centering
\input{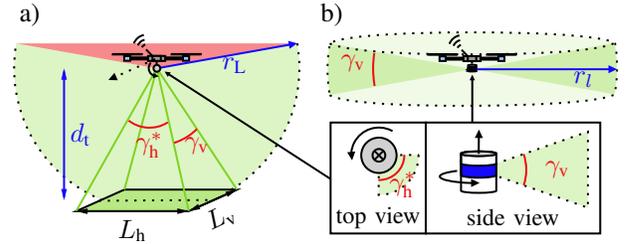}
\vspace{-0.25cm}
\caption{a) \ac{lidar} for vertical scanning; b) \ac{lidar} for horizontal scanning;}
\label{fig:LIDAR}
\end{figure}

\acp{lidar} measure distances by emitting laser pulses that reflect off objects and return to the receiver, allowing the system to calculate the distance based on the light's travel time. In combination with position and orientation data, the exact location of the point measurement is determined. By scanning across different directions, a precise 3D map of the environment, refastened by a point cloud, is created. The sensor's emitter and a receiver typically operate within a horizontal field of view \(\gamma_\text{h}\) and vertical field of view \(\gamma_\text{v}\) up to a defined effective \ac{lidar} range \(r_\text{L}\). Often both are mounted on a rotating axis to achieve a full scan of the horizontal plane \(\gamma_\text{h} = 2\,\pi\). The quality of these scans is expressed by the point cloud density \(R_\text{l}\) in points per square meters or by the spacing between points \(d_\text{p}\). Those depend on the vertical and horizontal angular resolutions, \(V_\text{res}\) and \(H_\text{res}\), as well as the distance to the target \(d_\text{t}\) \cite{Yang2023}. In the following, we derive constraints on the \ac{uav}'s operations for two different \ac{lidar} use cases, shown in Fig.~\ref{fig:LIDAR}, ensuring valid data collection. 

For the first \ac{lidar} configuration, the sensor is mounted on a gimbal to scan the vertical plane with its rotation axis parallel to the ground, as shown in Fig.~\ref{fig:LIDAR}a. This setup is typically used for ground measurements or mapping.
To ensure adequate point cloud density \(R_\text{L} = N_\text{p}/A_\text{L}\), the scanned area on the ground \(A_\text{L}=L_\text{v}\,L_\text{h} = 4\,d_\text{t}\,\tan\left(\gamma_\text{v}/2\right)\,\tan\left(\gamma_\text{h}^*/2\right)\) with the valid horizontal field of view \(\gamma_\text{h}^*\) and \(d_\text{t} = |z_\text{t}-z|\) has to be considered. The number of measurement points is then given by \(N_\text{p} = (\gamma_\text{v}\,\gamma_\text{h}^*)/(V_\text{res}\,H_\text{res})\). Ensuring a minimum point cloud density \(R_\text{L,min}\), the distance \(d_\text{t}\) must satisfy
\begin{align*}
d_\text{t} \leq \sqrt{\frac{\gamma_\text{v}\,\gamma_\text{h}^*}{4\,R_\text{L,min}\,\tan\left(\gamma_\text{v}/2\right)\,\tan\left(\gamma_\text{h}^*/2\right)}}, 
\end{align*}
while \(d_\text{t} \leq r_\text{L}\,\cos\left(\gamma_\text{v}/2\right)\) and \(d_\text{t} \leq r_\text{L}\,\cos\left(\gamma_\text{h}^*/2\right)\) ensure valid measurements.
Further, the \ac{uav}'s ground velocity \(v_\text{g}\) is constrained by
\begin{align*}
v_\text{g} \leq 2\,|z_\text{t}-z|\,\tan\left(\gamma_\text{v}/2\right)\,(1-\delta_\text{c})\,f_\text{L},
\end{align*}
where \(\delta_\text{L} \in [0,1]\) is the overlap of successive scans and \(f_\text{L}\) is the scanning rate. 

For the second \ac{lidar} configuration, the sensor is mounted to scan the horizontal plane with its rotation axis aligned with the \ac{uav}'s \(\text{z}^\text{B}\)-axis, as shown in Fig.~\ref{fig:LIDAR}b. This configuration is typically used for obstacle detection, requiring the \ac{uav} to limit its ground velocity so that it can react in time to detected obstacles. 
The ground velocity \(v_\text{g}\) must satisfy 
\begin{align*}
v_\text{g} \leq -a_\text{B,max}\,t_\text{R}+\sqrt{(a_\text{B,max}\,t_\text{R})^2+r_\text{L}},
\end{align*}
where \(t_\text{R}\) is the detection time and \(a_\text{B,max}\) is the maximum braking acceleration, approximated by \(a_\text{B,max} \approx \alpha_\text{max}\text{g}\). Additionally, the tilt angle \(\alpha\) should be constrained by the \ac{lidar}’s vertical field of view \( \gamma_\text{v} \) to maintain a clear detection of obstacles in front of the \ac{uav}, with 
\begin{align*}
\alpha \leq \gamma_\text{v}/2.
\end{align*}

\section{Energy Aware Multicopter Model}
\label{cha:model}
\begin{figure}[H]
\centering
\input{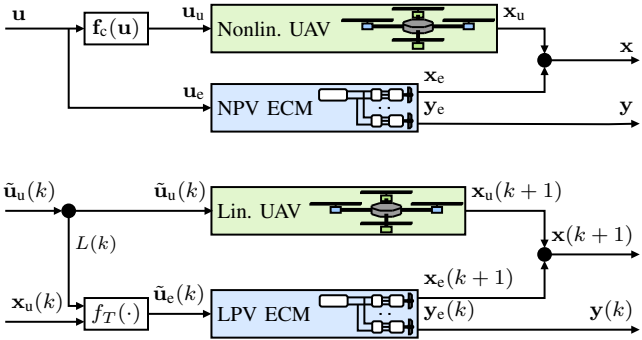}
\vspace{-0.25cm}
\caption{Structure of the nonlinear (top) and linear (bottom) energy aware multicopter model}
\label{fig:C_Setup}
\end{figure}

The derived multicopter and energy consumption models are combined, as illustrated in Fig.~\ref{fig:C_Setup}, to obtain the discrete-time \ac{npv} energy-aware multicopter model
\begin{equation*}
\label{eqn:NPV_EA_UAV}
\begin{aligned} 
\dot{\mathbf{x}} &= \mathbf{f}(\mathbf{x},\mathbf{u}) + \boldsymbol{\Gamma}_\text{x} = 
\begin{pmatrix}    \mathbf{f}_\text{u}\left(\mathbf{x}_\text{u},\mathbf{f}_\text{c}(\mathbf{u})\right) \\
\mathbf{f}_\text{e}\left(\mathbf{x}_\text{e},\mathbf{u}\right)\end{pmatrix} + 
\begin{pmatrix} \boldsymbol{\Gamma}_\text{u,x} \\ \boldsymbol{\Gamma}_\text{e,x}\end{pmatrix},& \\   
\mathbf{y} &= \mathbf{g}(\mathbf{x},\mathbf{u}) + \boldsymbol{\Gamma}_\text{y} = \mathbf{g}_\text{e}(\mathbf{x}_\text{e},\mathbf{u}) + \boldsymbol{\Gamma}_\text{e,y},
\end{aligned}
\end{equation*}
with the state \(\mathbf{x} = (\mathbf{x}_\text{u}^\top, \mathbf{x}_\text{e}^\top)^\top\), the input \(\mathbf{u} = (\Omega_1,\dots, \Omega_{N_\text{M}})^\top\) and the output \(\mathbf{y} = \mathbf{y}_\text{e}\). 
The transformation \(\mathbf{u}_\text{u} = \mathbf{f}_\text{c}(\mathbf{u})\) of the input \(\mathbf{u}\) into the input \(\mathbf{u}_\text{u}\) of the nonlinear multicopter model \eqref{eqn:Nonlin_Multicopter} is derived in \eqref{eqn:T} and \eqref{eqn:tau}.

Meanwhile, the discrete-time \ac{lpv} energy aware multicopter model with the reduced input \(\tilde{\mathbf{u}} = (\tilde{\mathbf{u}}_\text{u}^\top,\tilde{\mathbf{u}}_\text{e}^\top)^\top\) is given by
\begin{equation*}
\label{eqn:LPV_EA_UAV}
\begin{aligned}
\mathbf{x}(k+1) &= 
\mathbf{A}_\text{d}\mathbf{x}(k) + 
\mathbf{B}_\text{d}\tilde{\mathbf{u}}(k) + 
\mathbf{E}_\text{d} + \boldsymbol{\Gamma}_\text{x}, \\
\mathbf{y}(k) &= 
\mathbf{C}_\text{d}\mathbf{x}(k) + 
\mathbf{D}_\text{d}\tilde{\mathbf{u}}(k) + \mathbf{y}_\text{SP} + \boldsymbol{\Gamma}_\text{y}. 
\end{aligned}
\end{equation*}
It comprises the discrete-time state-space matrices
\begin{align*}
    &\mathbf{A}_\text{d} = 
    \begin{bmatrix} 
    \mathbf{A}_{\text{d,u}} & \mathbf{0} \\ 
    \mathbf{0}& \mathbf{A}_{\text{d,e}}
    \end{bmatrix}, 
    &&\mathbf{B}_\text{d} = 
    \begin{bmatrix}
    \mathbf{B}_{\text{d,u}} & \mathbf{0}\\ 
    \mathbf{0} & \mathbf{B}_{\text{d,e}}    
    \end{bmatrix},  
    &&\mathbf{E}_\text{d} = 
    \begin{bmatrix}
    \mathbf{0} \\
    \mathbf{E}_{\text{d,e}}    
    \end{bmatrix},\\    
    &\mathbf{C}_\text{d} = 
    \begin{bmatrix}
    \mathbf{0} & \mathbf{C}_{\text{d,e}}
    \end{bmatrix},  
    &&\mathbf{D}_\text{d} = 
    \begin{bmatrix}
    \mathbf{0} & \mathbf{D}_{\text{d,e}}    
    \end{bmatrix}.
    &&  
\end{align*}
\vspace{-0.5cm}

The separation of the inputs for the multicopter model and the \ac{ecm} allows for considering aerodynamic effects that are not included in the linear multicopter model. Since the horizontal and vertical motion dynamics are not coupled, we correct the necessary thrust during tilted flight by
\begin{align}
\label{constr:thrust}
    \Delta T = \sqrt{T_\text{v}^2+T_\text{h}^2} - T_\eta - T_\text{SP}.
\end{align}
Here, \( T_\text{v} = T\,\cos(\alpha) = T_\text{SP} + L\) is the vertical thrust component and \(T_\text{h} = T\,\sin(\alpha) \approx m\,\text{g}\,\alpha\) is the horizontal thrust component.
It is observed that a small-size multicopter in forward flight consumes equal or even less power compared to hovering until a threshold velocity is reached \cite{Theys2020,Cabuk2024}. This is the result of multiple rotor efficiency increasing aerodynamic effects, such as the increased air inflow velocity through the rotors or the effective translational lift \cite{Wall2020,HH2019}. 
Due to the nonlinear and difficult-to-model nature of these effects, we approximate the increased efficiency by reducing the necessary thrust for velocities below the threshold \(v_\text{th}\) by 
\begin{align*}
    T_\eta = 
    \begin{cases}
        \frac{\eta_\text{ETL}\,m\,\text{g}}{v_\text{th}}\,v_\text{g} & \text{for} \ v_\text{g} \leq v_\text{th},\\ 
        \eta_\text{ETL}\,m\,\text{g} & \text{for} \ v_\text{g} \geq v_\text{th}.
    \end{cases}
\end{align*}
Here, \(\eta_\text{ETL} = \sqrt{1 + \left(\frac{c_\text{F}\,v_\text{th}}{m\,\text{g}}\right)^2}-1\) results in an equal power consumption for hovering and for steady horizontal flight with \(v_\text{g}=v_\text{th}\).

Accurately representing the vehicle's capabilities, the constraints outlined in Sections \ref{sec:UAV_constr} and \ref{sec:ECM_constr} must be satisfied. Further, if the \ac{uav} is equipped with a sensor the constraints in Section \ref{cha:sensor} must be considered during measurements.

\begin{remark_new}
    For the parameter identification the masses of the \ac{uav}, battery and equipment has to be summed up to the total mass \(m\).
    Additionally, the inertia tensor \(\mathbf{J}\) must be adjusted regarding the battery and equipment, preferring to be attached close to the \ac{uav}'s body. 
\end{remark_new}

\section{Discussion}
\label{cha:Discussion}
In the following, we discuss the modeling of the \acp{uav} and their energy consumption,and validate them by actual test flight data.  

\subsection{Multicopter Model}
In Section \ref{cha:UAVModel}, we derived a multicopter model \eqref{eqn:Nonlin_Multicopter} that is suitable for \ac{uas} swarms with various \acp{uav} because it is easily adaptable to fit all sorts of multicopter configurations. However, the models have their limitations in terms of accuracy. We employ Assumption \ref{ass:aero_const} for the aerodynamic parameters \(k_\text{F}\) and \(k_\text{M}\), which we define for the hovering state (no external airflow, constant motor speeds). This leads to an estimation error for the aerodynamic forces and torques. In addition, the models have no upper limit for their flight altitude, since the aerodynamic parameters do not decrease depending on the air pressure. Additional aerodynamic effects such as the effective translational lift or the dynamic air inflow of the rotors, should be considered to improve the accuracy \cite{Wall2020}. The approximations on the dynamics and parameters result in uncertainties, which should be considered for path planning to ensure that the paths are safe and will not lead to collisions. Unfortunately, these are impossible to predict and only can be partly modeled.
Furthermore, the mapping between the inputs of the linear multicopter model \eqref{eqn:Lin_Multicopter} to the motor speeds is only valid under the condition outlined in Remark \ref{remark:psi}.

\subsection{Energy Consumption Models}
In Section \ref{cha:ECM}, we derived an \ac{ecm} \eqref{eqn:Nonlin_ECM} that is easily adaptable to different electric-propelled \acp{uav}. To validate the \ac{ecm} and the individual component models, we identify the model parameters of a "Holybro S500 V2", which are listed in Appendix \ref{cha:AppendixA}. Some parameters are taken directly from data sheets, other unknown parameters are fitted using the \textit{greyest} algorithm in Matlab, see \cite{Mathworks(2023)}, and data from a calibration flight (TD), recorded by the Pixhawk autopilot \citep{Pixhawk(2023)}. The models are validated with measurements from additional test flights, which are defined in Appendix \ref{cha:AppendixB}.
\vspace{-0.25cm}
\begin{figure}[H]
\centering
\input{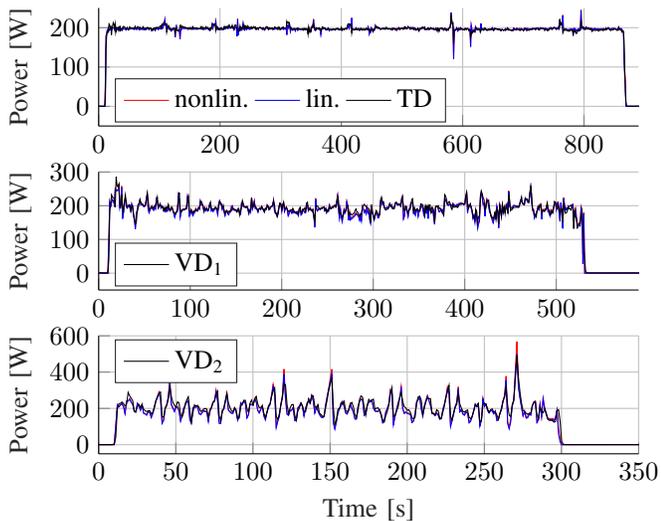}
\vspace{-0.75cm}
\caption{Supply Power estimation of the nonlinear (red) and linear (blue) \acs{esc-bldc} models, compared to measurements (black) of actual test flights}
\label{fig:ESC_BLDC_Validation}
\end{figure}
\vspace{-0.25cm}
Fig.~\ref{fig:ESC_BLDC_Validation} shows the combined supply power estimation of all four \ac{bldc} motors for the nonlinear and linearized \acs{esc-bldc} model. It turns out, that the supply power is estimated adequately during flight. However, the nonlinear model overestimates the supply power at high motor speeds. This behavior results from Assumption \ref{ass:aero_const} in \eqref{eqn:BLDC_P1}, since the supply power increases with the fourth power of \(\Omega\), which is actual damped by a decreasing \(k_\text{M}\). It also shows that the linearized model often underestimates the power consumption slightly for motor speeds below the set point.
\vspace{-0.25cm}
\begin{figure}[H]
\centering
\input{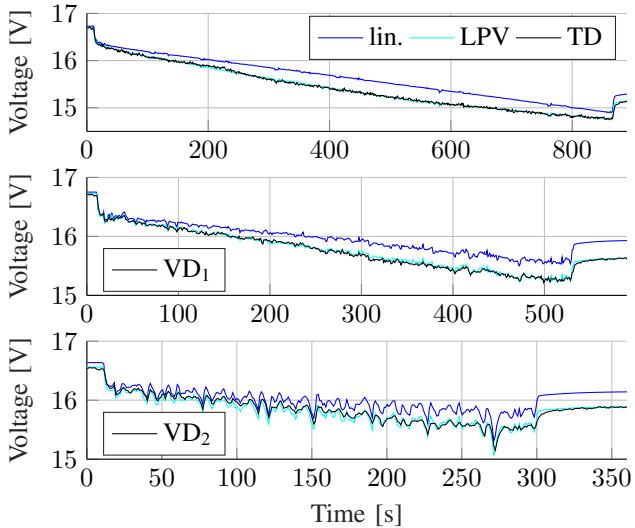}
\vspace{-0.25cm}
\caption{Battery voltage estimation of the linear (blue) and \ac{lpv} (cyan) battery models, compared to measurements (black) of actual test flights}
\label{fig:Bat_Validation}
\end{figure}
\vspace{-0.25cm}
Fig.~\ref{fig:Bat_Validation} shows the battery voltage estimation of the battery model \eqref{eqn:Bat} employing a linear and a \ac{lpv} Thevenin model as \ac{lib} cells. The \ac{lpv} battery model estimates the battery voltage with high accuracy. However, the estimation error increases during the relaxation phase. This is a known characteristic of the Thevenin model and is solved by the Dual-Polarization Model, which uses an additional RC-network in series to represent different polarization effects \citep{He(2011),Nikolian(2014)}. 
\vspace{-0.25cm}
\begin{figure}[H]
\centering
\input{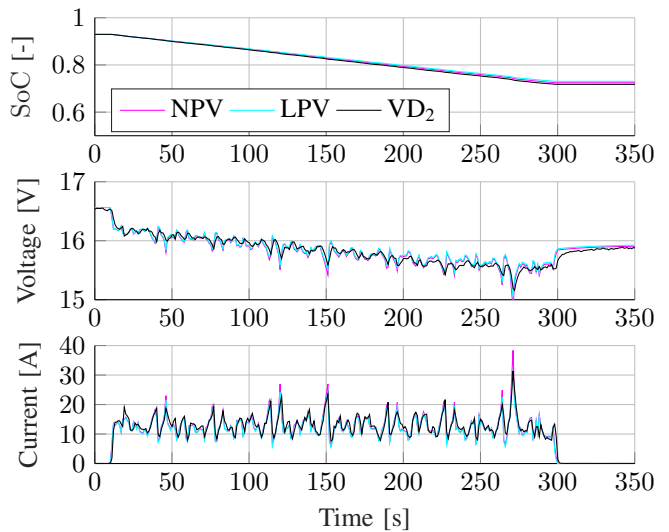}
\vspace{-0.75cm}
\caption{Energy consumption estimation of the \ac{npv} (magenta) and \ac{lpv} (cyan) \acp{ecm}, compared to measurements (black) of validation test flight VD\(_2\)}
\label{fig:ECM_Validation_VD_2}
\end{figure}
\vspace{-0.25cm}
To validate the combined \ac{ecm}, the energy consumption estimates of several \ac{ecm} formulations are compared with the measurements from the test flights. As an example, Fig.~\ref{fig:ECM_Validation_VD_2} shows the battery state measurements and estimates using the \ac{npv} and \ac{lpv} \acp{ecm} for the second validation test flight. 
While both models slightly underestimate the energy consumption, the \ac{npv} \acp{ecm} show better estimation results than the \ac{lpv} \acp{ecm} due to the overestimated supply power of the \acs{esc-bldc} model. All \acp{ecm} show the same estimation errors for the relaxing phase as their corresponding battery models.
As metric for comparison, we use the state of charge estimation error \(f_\text{error} = |\Delta\mathrm{SoC}-\Delta\tilde{\mathrm{SoC}}| \cdot 100\%,\) where \(\Delta\mathrm{SoC}\) and \(\Delta\tilde{\mathrm{SoC}}\) are the actual and estimated differences in the state of charge change \(\mathrm{SoC}\) after five minutes of flight.
\vspace{-0.25cm}
\begin{table}[H]
\caption{State of charge estimation errors after five minutes.}
\begin{center}
\begin{tabular}{c|cccc}
     & nonlin. & NPV & lin. & LPV \\ \hline
TD   & 0.40\% & 0.22\% & 0.46\% & 0.29\%\\     
VD\(_1\) & 0.34\% & 0.18\% & 0.55\% & 0.39\%\\     
VD\(_2\) & 0.89\% & 0.66\% & 1.46\% & 1.25\%
\end{tabular}
\end{center}
\label{tab:SOC_error}
\end{table}
\vspace{-0.25cm}
As shown in Tab.~\ref{tab:SOC_error}, the \acp{ecm} can accurately estimate the state of charge, and the estimation errors are less than 2\% even for flight profiles that deviate significantly from the set point. Furthermore, parameter-varying variants of the \acp{ecm} can improve the estimation accuracy. The \acp{ecm} show a self-amplifying effect for the estimation errors, which is present in the simulation because we use do not correct the state estimations by actual measurements. 

\subsection{Energy Aware Multicopter Model in Path Planning}
\begin{figure}[H]
    \centering
    \includegraphics[width=0.5\textwidth]{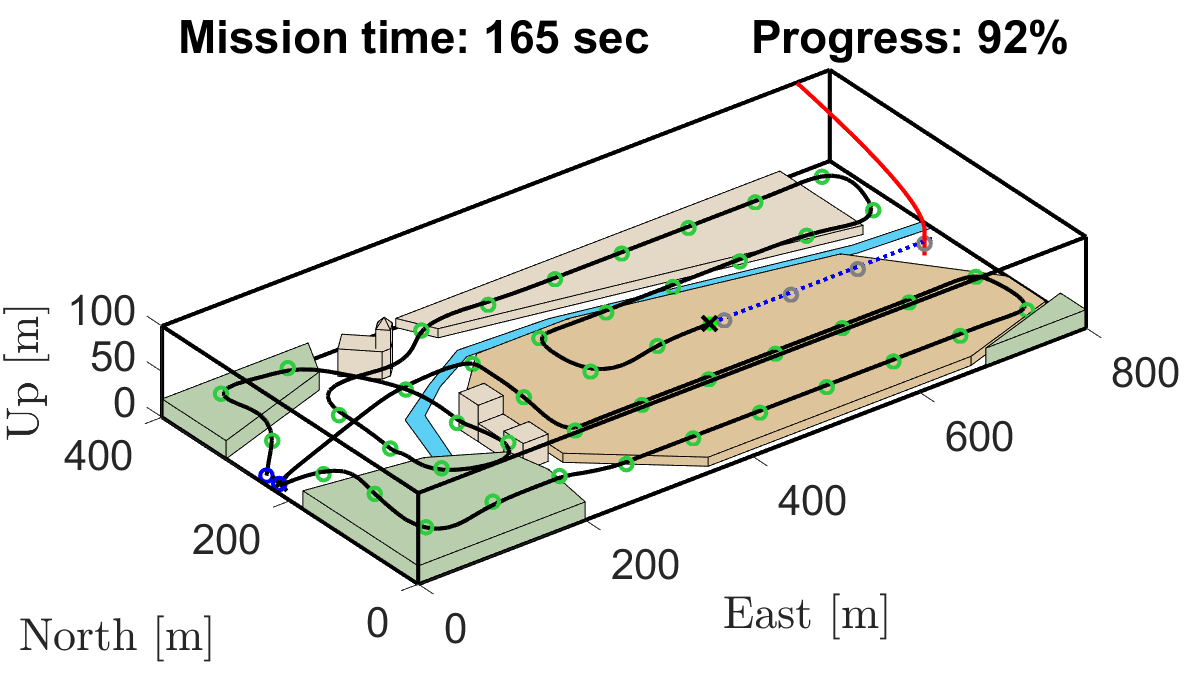}
    \vspace{-0.75cm}
    \caption{\ac{uas} swarm on a search-and-rescue mission in a small flooded village employing two \acp{uav} to autonomously cover the search area; \linebreak \ac{uav}: position (black cross), past path (black line), planned path (blue dots); \linebreak
    Moving obstacle (e.g. a rescue helicopter): past path (red line) \cite{Gasche2024_PPA}}
    \label{fig:Sim_2}
\end{figure}

\begin{figure}[H]
    \centering
    \input{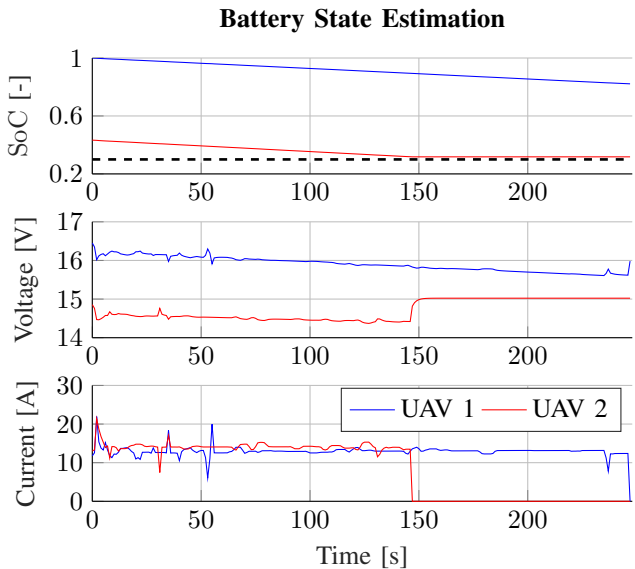}
    \vspace{-0.3cm}
    \caption{Energy consumption simulation for two \acp{uav}  \cite{Gasche2024_PPA}}
    \label{fig:Sim_3}
\end{figure}

In \cite{Gasche2024_PPA}, the developed discrete-time linear energy aware multicopter model and the parameters of the "Holybro S500 V2", are implemented in an online moving horizon path planning algorithm to plan energy efficient paths for a \ac{uas} swarm on a search and rescue mission, resulting in the flight paths drawn in  Fig.~\ref{fig:Sim_2}. Additionally, Fig.~\ref{fig:Sim_3} illustrates the results of the energy consumption simulation of the two \acp{uav}. Note, that \ac{uav} 2 is already partly discharged at the beginning of the mission and returns during the mission to its base, where it lands and deactivates itself.
\vspace{-0.5cm}
The state of charge, the battery voltage and the battery current of \ac{uav} 1 and \ac{uav} 2 are shown in blue and red, respectively. It can be seen that \ac{uav} 2 is drawing a higher current to compensate for its lower charge and battery voltage. The \ac{ecm} can reproduce the desired effects that affect the energy consumption. It increases when climbing, turning, braking, or accelerating and decreases when descending. Due to \eqref{constr:thrust} the power consumption during horizontal flight at maximum velocity is only slightly higher, compared to a steady hover flight, which also is observable in reality. 

\section{Conclusion}
\label{cha:Conclusion}
In this study, we developed a modular modeling approach for multicopter \acp{uav}, with a focus on their energy consumption and applicability to control and planning applications. 

We derived an adaptable multicopter model capable of representing a wide range of multicopter configurations, making it suitable for heterogeneous \ac{uas} swarms. In doing so, we incorporated important factors such as vehicle capabilities, external disturbances, and uncertainties to ensure the model closely represents real-world flight dynamics. 
Moreover, we derived an \ac{ecm} for electric-propelled \acp{uav}. By adopting a component-based modeling approach, the detailed model design enables us to consider different \ac{uav} designs, enhancing the adaptability of the model. A scheme for an electric-propelled \ac{uav}'s power train was designed, consisting of \ac{bldc} motors (modeled as simplified direct current motors), \acp{esc} (modeled as DC-DC-converters), and \acp{lib} (modeled as networks of \ac{lib} cells employing an equivalent circuit design approach). Together, these components form the \ac{ecm}, which allows for accurate estimation of the battery state based on motor speeds and power demands from additional subsystems, like the avionics or the payload. This approach ensures adaptability across a wide range of electric-propelled \acp{uav}. 
The combined nonlinear model was further linearized and discretized for a multicopter\ac{uav}, making it suitable for integration into control and planning algorithms. This is demonstrated by its use in the path planner developed in the companion study. 
In addition, sensor models for camera and \ac{lidar} systems were derived, offering the flexibility to integrate mission-specific constraints and sensor limitations into the planning process. 

Overall, this modular approach provides a solid foundation for detailed analysis and optimization of multicopter \ac{uav} operations, enabling more precise control and mission execution under real-world conditions.

\appendices

\section{Discrete-time Linear State Space Models}
\label{cha:AppendixD}
We linerize the nonlinear models \eqref{eqn:Nonlin_Multicopter} and \eqref{eqn:Nonlin_ECM} around the hovering state with a fully charged battery
\begin{alignat*}{3}
\mathbf{x}_\text{u,SP} &= (0,\dots,0)^\top,  &\mathbf{x}_\text{e,SP} &= (0,0)^\top, \\
\mathbf{u}_\text{u,SP} &= (T_\text{SP},0,0,0,0)^\top, \quad
&\mathbf{u}_\text{e,SP} &= (\Omega_\text{SP},\dots, \Omega_\text{SP})^\top,\\
\mathbf{d}_\text{u,SP} &= (0,\dots,0)^\top, & 
\end{alignat*}
wherein the multicopter maintains its position while the thrust \(T_\text{SP}=m\,\text{g}\) compensates for the weight force, resulting in motor speeds of \(\Omega_\text{SP}=\sqrt{\frac{m\,\text{g}}{N_\text{M}\,k_\text{F}}}\). In the following uncertainties are not considered. 
\begin{remark_new}
In linearized models, deviations from the set point are used. For example, the state deviation \(\Delta\mathbf{x}\) is defined as \( \Delta\mathbf{x} = \mathbf{x} - \mathbf{x}_\text{SP}\). However, when the variable is 0 at the set point, \(\Delta\) is omitted for clarity.
\end{remark_new}

\subsection{Discrete-time Linear Multicopter Model}
For the multicopter model, this set point is an equilibrium point and the linear model is given by 
\begin{align*}
\dot{\mathbf{x}}_\text{u} &= \mathbf{f}_\text{u,lin}(\mathbf{x}_\text{u},\tilde{\mathbf{u}}_\text{u},\mathbf{d}_\text{u}) \\ & \\ &= 
\begin{pmatrix}
v_\text{x}\\
v_\text{y}\\
v_\text{z}\\
-\text{g}\,(\theta\,\cos(\psi_\text{SP}) + \phi\,\sin(\psi_\text{SP})) - c_\text{Fx}\,(v_\text{x} - v_\text{w,x})/m \\
-\text{g}\,(\theta\,\sin(\psi_\text{SP}) - \phi\,\cos(\psi_\text{SP})) - c_\text{Fy}\,(v_\text{y} - v_\text{w,y})/m \\
-L/m - c_\text{Fz}\,(v_\text{z} - v_\text{w,z})/m \\
\omega_\text{x}\\
\omega_\text{y}\\
\omega_\text{z}\\
(\tau_\text{x}-\omega_\text{x}\,c_{\tau\text{x}})/J_\text{xx}\\
(\tau_\text{y}-\omega_\text{y}\,c_{\tau\text{y}})/J_\text{yy}\\
(\tau_\text{z}-\omega_\text{z}\,c_{\tau\text{z}})/J_\text{zz}
\end{pmatrix}.
\end{align*} 

\noindent
For \(\psi_\text{SP}=0\), the transformation matrices \(\mathbf{R}_\text{B}^\text{I}\) and \(\mathbf{R}_\Psi^{-1}\) equal identity matrices, resulting in a direct transformation of vectors and rotation rates between the body-fixed and inertial frame. Therefore, the alignment of the \(\text{x}^\text{B}\)- and \(\text{y}^\text{B}\)-axis is not affected by the yaw angle \(\psi\) anymore. Likewise, the alignment of the \(\text{z}^\text{B}\)-axis is not affected by the roll and pitch angels \(\phi, \theta\). Since the thrust \(T\) now only acts in negative \(\text{z}^\text{I}\)-direction, we replace it with the lift \(L \approx T - T_\text{SP}\). Further, the difference in rotor speeds \(\Omega_\text{r}\), defined by \eqref{eqn:Omega_r}, does not affect the dynamics anymore, so we reduce the input to
\begin{equation*} 
\tilde{\mathbf{u}}_\text{u} = (L,\tau_\text{x},\tau_\text{y},\tau_\text{z})^\top.
\end{equation*}
This linear model is discretized with a sampling time of \(\Delta t\) using Taylor-Lie series, as described in Appendix \ref{cha:AppendixC}. In order to more accurately approximate the discrete-time model a discretization order \(N_\text{dis} \geq 2\) is recommended due to the high dynamics of the multicopter. Finally, the discrete-time linear state-space multicopter model is given by
\begin{equation*}
\begin{aligned}
    \mathbf{x}_\text{u}(k + 1) = \mathbf{A}_\text{d,u}\,\mathbf{x}_\text{u}(k)+\mathbf{B}_\text{d,u}\,\tilde{\mathbf{u}}_\text{u}(k) + \mathbf{H}_\text{d,u}\,\mathbf{d}_\text{u}(k),
\end{aligned}
\end{equation*}
where \(\mathbf{A}_\text{d,u}\), \(\mathbf{B}_\text{d,u}\) and \(\mathbf{H}_\text{d,u}\)  are the discrete-time system, input and disturbance matrices.

\subsection{Discrete-time Linear Energy Consumption Model}
We start with the already linear \ac{lib} equations \eqref{eqn:Bat}
\begin{align*}
    \dot{\mathbf{x}}_\text{b} &= \mathbf{A}_\text{b}\,\mathbf{x}_\text{b} + \mathbf{B}_\text{b}\,\mathbf{u}_\text{b},\\
    \mathbf{y}_\text{b} &= \mathbf{C}_\text{b}\,\mathbf{x}_\text{b} + \mathbf{D}_\text{b}\,\mathbf{u}_\text{b},
\end{align*}
with the battery state \(\mathbf{x}_\text{b} = (\mathrm{DoD},u_\text{th})^\top\), input \(\mathbf{u}_\text{b} = i_\text{b}\) and output \(\mathbf{y}_\text{b} = (\mathrm{SoC},u_\text{b},i_\text{b})^\top,\) and the state-space matrices 
\begin{alignat*}{3}
\mathbf{A}_\text{b} &= \begin{bmatrix} 0 & 0 \\ 0 & -1/(R_\text{th}\,C_\text{th})\end{bmatrix}, \quad
&\mathbf{B}_\text{b} = \begin{bmatrix} \eta_\text{b}/Q_\text{b} \\ 1/(N_\text{P}\,C_\text{th})\end{bmatrix},\\
\mathbf{C}_\text{b} &= \begin{bmatrix} -1 & 0 \\ N_\text{S}\,b_1 & -N_\text{S} \\ 0 & 0 \end{bmatrix}, \quad 
&\mathbf{D}_\text{b} = \begin{bmatrix} 0 \\ -(N_\text{S}\,R_\text{int})/N_\text{P} \\ 1 \end{bmatrix}.
\end{alignat*}

Next, we derive from \eqref{eqn:BLDC_P1} a linear approximation of the power consumption of a \ac{bldc} motor 
\begin{align*}
    \Delta p_\text{DC} &= \frac{\text{d}p_\text{DC}}{\text{d}\Omega^2}\bigg|_\text{SP}\,\Delta\Omega^2 = \kappa_\text{DC}\,\Delta\Omega^2
\end{align*}
depending on the motor speed squared around the set point, where \(\kappa_\text{DC}\) substitutes for
\begin{equation*}
    \kappa_\text{DC} = K_\text{V}^2\,R_\text{DC}(2\,k_\text{M}^2\,\Omega_\text{SP}^2+3\,k_\text{M}\,D_\text{f}\,\Omega_\text{SP}+D_\text{f}^2)+\frac{3}{2}\,k_\text{M}\,\Omega_\text{SP} + D_\text{f}.
\end{equation*}
Combining all \ac{bldc} motors together, while considering \eqref{eqn:T} and Assumption \ref{ass:UAV_2}, we formulate the total power consumption of the \ac{bldc} motors \(p_{\text{DC},\Sigma}\) depending on the thrust \(T\) by 
\begin{equation*}
     \Delta p_{\Sigma} = \sum_{i=1}^{N_\text{M}} \Delta p_\text{DC} = \frac{\kappa_\text{DC}}{k_\text{F}}\,\Delta T, 
\end{equation*} 
The power consumption of the \ac{bldc} motors during hovering 
\begin{align*}
    p_\text{DC,SP} &=  R_\text{DC}\,K_\text{V}^2\,\left(D_\text{f}\,\Omega_\text{SP} + k_\text{M}\,\Omega_\text{SP}^2\right)^2 + D_\text{f}\,\Omega_\text{SP}^2 + k_\text{M}\,\Omega_\text{SP}^3,\\
    p_{\Sigma\text{,SP}} &= N_\text{M}\,p_\text{DC,SP}
\end{align*} 
is calculated by \eqref{eqn:BLDC_P1}.

To connect the \ac{bldc} motors with the battery, we derive from \eqref{eqn:ib} the linear approximation for the current \(i_\text{b}\) drawn from the battery by the \acp{esc}:
\begin{align*}
    \Delta i_\text{b} = \sum_{i=1}^{N_\text{M}} \Delta i_{\text{ESC},i} = \kappa_{\mathrm{DoD}}\mathrm{DoD} + \kappa_\text{uth}u_\text{th} + \kappa_\text{p}\Delta p_\Sigma,
\end{align*} 
where \(\kappa_{\mathrm{DoD}}\),  \(\kappa_\text{uth}\) and \(\kappa_\text{p}\) substitute for 
\begin{alignat*}{3}
    \kappa_{\mathrm{DoD}} &= \frac{N_\text{P}\,b_1}{2\,R_\text{int}}(1-\kappa), \quad
    &\kappa_\text{uth} &= \frac{N_\text{P}}{2\,R_\text{int}}(\kappa-1),\\
    \kappa_\text{p} &= \frac{\kappa}{N_\text{S}\,b_0\,\eta_\text{ESC}}, \quad 
    &\kappa^{-1} &= \sqrt{1-\frac{4\,R_\text{int}}{N_\text{S}\,N_\text{P}\,b_0^2\,\eta_\text{ESC}}p_{\Sigma\text{,SP}}}.
\end{alignat*} 
From \eqref{eqn:BLDC_P1} and \eqref{eqn:ib}, we derive the current draw from the battery during hovering 
\begin{align*}
    i_\text{b,SP} =  \frac{N_\text{P}\,b_0}{2\,R_\text{int}}\left(1-\kappa^{-1}\right).
\end{align*} 

Finally, we define for the linear multicopter \ac{ecm}, the state \(\mathbf{x}_\text{e} = \mathbf{x}_\text{b} (\mathrm{DoD},u_\text{th})^\top\), the input \(\tilde{\mathbf{u}}_\text{e} = \Delta T\) and the output \linebreak \(\mathbf{y}_\text{e} = \mathbf{y}_\text{b} = (\mathrm{SoC},u_\text{b},i_\text{b})^\top\), which is  \(\mathbf{y}_\text{e,SP} = \mathbf{D}_\text{b}\,i_\text{b,SP}\) at the set point. Combining all components of the power train, the linear \ac{ecm} is given by 
\begin{alignat*}{3}
    &\dot{\mathbf{x}}_\text{e} &= \mathbf{f}_\text{e,lin}(\mathbf{x}_\text{e},\tilde{\mathbf{u}}_\text{e}) &= \mathbf{A}_\text{e}\,\mathbf{x}_\text{e} + \mathbf{B}_\text{e}\,\tilde{\mathbf{u}}_\text{e} + \mathbf{E}_\text{e},\\
    &\mathbf{y}_\text{e} &= \mathbf{g}_\text{e,lin}(\mathbf{x}_\text{e},\tilde{\mathbf{u}}_\text{e}) &= \mathbf{C}_\text{e}\,\mathbf{x}_\text{e} + \mathbf{D}_\text{e}\,\tilde{\mathbf{u}}_\text{e} + \mathbf{y}_\text{e,SP},
\end{alignat*}
with the state-space matrices
\begin{alignat*}{3}
 \mathbf{A}_\text{e} &= \mathbf{A}_\text{b} + \mathbf{B}_\text{b}\begin{pmatrix} \kappa_{\mathrm{DoD}} & \kappa_\text{uth}\end{pmatrix}, \quad
 &\mathbf{B}_\text{e} &= \mathbf{B}_\text{b}\,\frac{\kappa_\text{p}\,\kappa_\text{DC}}{k_\text{F}},\\
 \mathbf{C}_\text{e} &= \mathbf{C}_\text{b} + \mathbf{D}_\text{b}\begin{pmatrix} \kappa_{\mathrm{DoD}} & \kappa_\text{uth}\end{pmatrix}, \quad
 &\mathbf{D}_\text{e} &= \mathbf{D}_\text{b}\,\frac{\kappa_\text{p}\,\kappa_\text{DC}}{k_\text{F}}.
 \end{alignat*}
Since the set point is not an equilibrium point, we add \[\mathbf{E}_\text{e} = \mathbf{B}_\text{b}\,i_\text{b,SP}\] as an offset for the energy consumption during hovering.

This linear model is discretized with a sampling time of \(\Delta t\) using Taylor series, since a discretization order \(N_\text{dis} = 1\) is sufficient due to the direct influence of the input on the state. Then, the discrete-time linear state-space \ac{ecm} is given by
\begin{equation*}
\begin{aligned}
    \mathbf{x}_\text{e}(k + 1) &= \mathbf{A}_\text{d,e}\,\mathbf{x}_\text{e}(k)+\mathbf{B}_\text{d,e}\,\tilde{\mathbf{u}}_\text{e}(k) + \mathbf{E}_\text{d,e}, \\ 
    \mathbf{y}_\text{e}(k) &= \mathbf{C}_\text{d,e}\,\mathbf{x}_\text{e}(k)+\mathbf{D}_\text{d,e}\,\tilde{\mathbf{u}}_\text{e}(k) + \mathbf{y}_\text{e,SP},
\end{aligned}
\end{equation*}
with the discrete-time system, input and disturbance matrices 
\begin{alignat*}{3}
 \mathbf{A}_\text{d,e} &= \mathbf{I}+\mathbf{A}_\text{e}\Delta t, \quad
 &\mathbf{B}_\text{d,e} &= \mathbf{B}_\text{e}, \quad \mathbf{E}_\text{d,e} &= \mathbf{E}_\text{e}\Delta t,\\
 \mathbf{C}_\text{d,e} &= \mathbf{C}_\text{e}, \quad &\mathbf{D}_\text{d,e} &= \mathbf{D}_\text{e}. &
 \end{alignat*}
\begin{remark_new}
    Here, the parameters \(b_0\) and \(b_1\) can be ether selected according to \eqref{eqn:b0b1_lin} or \eqref{eqn:b0b1_LPV} for the linear or \ac{lpv} \ac{ecm}. 
\end{remark_new}

\section{Discretization Using Taylor-Lie Series}
\label{cha:AppendixC}
For discretizations, we adopt Lie-derivatives 
\begin{equation*}
\begin{split}
    L_\mathbf{f}^1\,\mathbf{f}(t,\mathbf{x},\mathbf{u},\mathbf{d}) = &\frac{\partial\,\mathbf{f}}{\partial\,t}(t,\mathbf{x},\mathbf{u},\mathbf{d}) \dots\\ &+ \nabla_\mathbf{x}\,\mathbf{f}(t,\mathbf{x},\mathbf{u},\mathbf{d}) \cdot \mathbf{f}(t,\mathbf{x},\mathbf{u},\mathbf{d}), \\
    L_\mathbf{f}^k\,\mathbf{f}(t,\mathbf{x},\mathbf{u},\mathbf{d}) = &\frac{\partial\left(L_\mathbf{f}^{k-1}\,\mathbf{f}\right)}{\partial\,t}(t,\mathbf{x},\mathbf{u},\mathbf{d}) \dots \\ &+ \nabla_\mathbf{x}\left(L_\mathbf{f}^{k-1}\,\mathbf{f}\right)(t,\mathbf{x},\mathbf{u},\mathbf{d}) \cdot \mathbf{f}(t,\mathbf{x},\mathbf{u},\mathbf{d}), 
\end{split}
\end{equation*}
to increase the accuracy of the discrete-time system dynamics approximation. Then, the discrete-time model is described by
\begin{equation*}
\begin{split}
    \mathbf{x}(k+1) = \mathbf{f}_\text{d}(t,\mathbf{x},\mathbf{u},\mathbf{d},\Delta t) &= \mathbf{x} + \mathbf{f}(t,\mathbf{x},\mathbf{u},\mathbf{d}) \cdot \Delta t \\ &+ \sum_{k=2}^{N_\text{dis}} L_\mathbf{f}^{k-1}\,\mathbf{f}(t,\mathbf{x},\mathbf{u},\mathbf{d}) \cdot \frac{\Delta t^k}{k!},
\end{split}
\end{equation*}
where \(\Delta t\) is a constant sampling time, and \(N_\text{dis}\) is the discretization degree  \citep{Kazantzis(2005)}.

\section{Model Parameters}
\label{cha:AppendixA}
\begin{table}[H]
\caption{Vehicle parameters and limitations}
\begin{center}
\begin{tabular}{|l|l|l|l|}
\multicolumn{4}{c}{Quadcopter "Holybro S500 V2" (with battery)} \\ \hline
\(m = 1.45\,\text{kg}\) & \(\ell = 0.24\,\text{m}\) & \multicolumn{2}{l|}{\(J_\text{xx} = J_\text{yy} = 0.0158\,\text{kg}\,\text{m}^2\)} \\ \hline
\(N_\text{M} = 4\) & \(v_\text{th} = 10\,\text{m}/\text{s}\)  &  \multicolumn{2}{l|}{\(J_\text{zz} = 0.0252\,\text{kg}\,\text{m}^2\)} \\ \hline
\(v_\text{max} = 13.5 \,\text{m}/\text{s}\) & \(v_\text{z,max} = 5 \,\text{m}/\text{s}\) & \multicolumn{2}{l|}{\(\delta_\text{UAS} = 0.37 \text{m}\)} \\ \hline
\(\alpha_\text{max} = 30^\circ\) &  \(\omega_\text{max} = 15^\circ/\text{s}\) & \multicolumn{2}{l|}{\(c_{\text{F},i} = 0.27\,(\text{N}\,\text{s})/\text{m}\)} \\ \hline
\multicolumn{2}{|l|}{\(l_{\text{x},i} = l_{\text{y},i} = \sqrt{2}/2\,\ell = 0.17\,\text{m}\)} & \multicolumn{2}{l|}{\(c_{\tau,i} = 0.1\,(\text{N}\,\text{m}\,\text{s})/\text{rad}\)} \\ \hline
\multicolumn{4}{c}{} \\ 
\end{tabular}
\begin{tabular}{|l|l|}
\multicolumn{2}{c}{\ac{esc} "BLHeli-S 20A" \& \ac{bldc} "AIR2216II" \& Rotor "T1045II"} \\ \hline
\(R_\text{DC} = 57.5\,\text{m}\Omega\) & \(k_\text{F} = 1.21 \cdot 10^{-5}\,\text{N}/(\text{rad}/\text{s})^2\) \\ \hline
\(K_\text{V} = 96.34\,(\text{rad}/\text{s})/\text{V}\)  & \(k_\text{M} = 1.74 \cdot 10^{-7}\,\text{Nm}/(\text{rad}/\text{s})^2\) \\ \hline
\(\Omega_\text{max} = 1032\,\text{rad}/\text{s}\) & \(J_\text{r} = 9.86 \cdot 10^{-5}\,\text{kg}\,\text{m}^2\) \\ \hline
\(u_\text{DC,norm} = 16\,\text{V}\) &  \(\eta_\text{ESC} = 0.86\) \\ \hline 
\multicolumn{2}{c}{} \\  
\end{tabular}
\begin{tabular}{|l|l|l|l|l|}
\multicolumn{4}{c}{\ac{lib} "Gens Ace B-50C-5000-4S1P-Bashing"} \\ \hline
\(N_\text{S} = 4\) & \(N_\text{P} = 1\) & \(Q_\text{b} = 18000\,\text{As}\) & \multicolumn{2}{l|}{\(\eta_\text{b} = 1\)} \\ \hline
\multicolumn{2}{|l|}{\(R_\text{int} = 6.62\,\text{m}\Omega\)} &  \(R_\text{th} = 1.56\,\text{m}\Omega\) &  \multicolumn{2}{l|}{\(C_\text{th} = 15.6\,\text{kF}\)} \\ \hline
\multicolumn{2}{|l|}{\(u_\text{c,min} = 2.75\,\text{A}\)} & \(u_\text{c,max} = 4.2\,\text{A}\) &   \multicolumn{2}{l|}{\(i_\text{discharge,max} = 250\,\text{A}\)} \\ \hline
\multicolumn{2}{|l|}{\(u_\text{b,norm} = 14.8\,\text{V}\)} & \(\mathrm{DoD}_\text{cutoff} = 0.85\) &   \multicolumn{2}{l|}{\(\mathrm{DoD}_\text{max} = 0.7\)} \\ \hline
\multicolumn{2}{|l|}{\(b_\text{0} = 4.2\,\text{V}\)} & \(b_\text{1} = -0.5765\,\text{V}\) &  \multicolumn{2}{l|}{\(\overline{\mathrm{DoD}}_0 = 0\)} \\ \hline
\multicolumn{2}{|l|}{\(b_\text{0,1} = 4.2\,\text{V}\)} & \(b_\text{1,1} = -0.8395\,\text{V}\) &  \multicolumn{2}{l|}{\(\overline{\mathrm{DoD}}_1 = 0.2\)} \\ \hline
\multicolumn{2}{|l|}{\(b_\text{0,2} = 4.1727\,\text{V}\)} & \(b_\text{1,2} = -0.7028\,\text{V}\) &   \multicolumn{2}{l|}{\(\overline{\mathrm{DoD}}_2 = 0.4\)} \\ \hline
\multicolumn{2}{|l|}{\(b_\text{0,3} = 4.0529\,\text{V}\)} & \(b_\text{1,3} = -0.4034\,\text{V}\) &  \multicolumn{2}{l|}{\(\overline{\mathrm{DoD}}_3 = 0.9\)} \\ \hline
\end{tabular}
\end{center}
\label{tab:1}
\end{table}   

\section{Test Flights with a "Holybro S500 V2"}
\label{cha:AppendixB}
\begin{itemize}
    \item Trainings Data (TD): hovering with frequently impulsive changes in altitude; indoors
    \item Validation Data 1 (VD\(_1\)): rectangle trajectory following; medium cruise velocity (\(v_\text{cruise} = 19\text{km}/\text{h}\)); outdoors
    \item Validation Data 2 (VD\(_2\)): rectangle trajectory following; high cruise velocity (\(v_\text{cruise} = 43\text{km}/\text{h}\)); outdoors
\end{itemize}
\begin{remark_new}
    The flight profile of the validation test flights does not represent a horizontal flight because the autopilot could not maintain altitude during braking or acceleration.
\end{remark_new}

\ifCLASSOPTIONcaptionsoff
  \newpage
\fi

\small
\bibliographystyle{plainnat}
\bibliography{paper}

\normalsize
\end{document}